\documentclass[acmsmall]{acmart}
\usepackage[utf8]{inputenc}
\usepackage[T1]{fontenc}
\usepackage{array}
\usepackage{booktabs}
\usepackage{multicol}
\usepackage{multirow}
\usepackage{tabularx}
\usepackage{longtable}
\usepackage{threeparttable}
\usepackage{enumitem} 
\usepackage{tcolorbox}
\usepackage[super]{nth}
\tcbuselibrary{breakable}
\usepackage{balance}
\usepackage{hyperref}
\usepackage{tikz}
\usepackage{color,soul}
\usepackage{amsmath}
\usepackage{algorithm}
\usepackage[noend]{algpseudocode}
\usepackage{balance}
\usepackage[toc,page]{appendix}
\usepackage{fixltx2e}
\usepackage{seqsplit}
\usepackage{xurl}

\usepackage{todonotes}

\newcommand{\ps}[1]

\newcommand{\sectopic}[1]{\vspace{0.2em}\par\noindent{\textit{\bfseries #1}}}

\begin{document}

\title{DRILS: Dependency-based Requirement Change Impact Analysis Using LLMs}

\author{Romina Etezadi}
\affiliation{\institution{School of Electrical Engineering and Computer Science, University of Ottawa}
 \city{Ottawa}
 \country{Canada}}
\email{retez068@uottawa.ca}

\author{Ikram Darif}
\affiliation{\institution{School of Electrical Engineering and Computer Science, University of Ottawa}
  \city{Ottawa}
  \country{Canada}}
\email{idarif@uottawa.ca}

\author{Lionel Briand}
\affiliation{\institution{School of Electrical Engineering and Computer Science, University of Ottawa}
  \city{Ottawa}
  \country{Canada}
  }
  \affiliation{\institution{Lero Centre, University of Limerick}
  \city{Limerick}
  \country{Ireland}
  }
\email{lbriand@uottawa.ca}

\author{Ramesh S}
\affiliation{\institution{General Motors, Detroit, Michigan}
  \city{Michigan}
  \country{United States}}
\email{ramesh.s@gm.com}

\author{Prakash Peranandam}
\affiliation{\institution{General Motors, Detroit, Michigan}
  \city{Michigan}
  \country{United States}}
\email{prakash.peranandam@gm.com}

\begin{abstract}
Throughout the life-cycle of a software system, change is an inevitable and recurring phenomenon. Requirements, far from being static artifacts, are equally susceptible to such changes, as they must continuously adapt to evolving stakeholder needs, emerging constraints or regulations, and shifting project scopes. Managing these changes effectively is therefore a fundamental challenge in software engineering. Given the limited resources available to requirements engineers, the high number of requirements, and the inherent complexity of Natural Language (NL) requirements, manually identifying the impact of a change request on requirements is both error-prone and time-consuming. This challenge is further compounded by the complex interdependencies between requirements. Large Language Models (LLMs), with their strong NL processing capabilities, offer promising solutions for requirements change impact analysis (CIA). However, their effectiveness, particularly when enhanced with advanced techniques, remains underexplored.

In this paper, we present \textit{DRILS}, an LLM-based agentic framework for requirements Change Impact Analysis (CIA). \textit{DRILS} employs a multi-agent architecture to generate candidate impact sets for a given change request, complemented by an LLM-as-a-judge to promote diversity during impact set selection. \textit{DRILS} also features two key components: RE-RAG, a carefully engineered RAG technique that provides focused domain context, and a requirement dependency graph (RDG) as an additional knowledge source. We empirically evaluate \textit{DRILS} across three datasets manually curated for the CIA task. Our evaluation demonstrates the usefulness of \textit{DRILS}'s individual steps, inputs, and outputs, as well as its superior performance compared with state-of-the-art baselines. Across datasets, \textit{DRILS} achieves an average recall of 87.1\%, precision of 76.3\%, and an $F_1$ score of 81\%. It outperforms all baselines across all evaluation metrics while achieving higher effectiveness at a lower cost. The largest relative gains are related to precision, with improvements of 41.1\%, 83.4\%, and 149.8\% over the closest baseline on each dataset. These gains are particularly significant for CIA, as higher precision reduces the number of false positives requiring manual validation. This substantially decreases the manual effort and time required for requirements CIA. Finally, we contribute three manually annotated datasets of requirements change impact sets to support future research.

\end{abstract}

\maketitle

\section{Introduction}\label{sec:introduction}
Requirements form the foundation of system development and maintenance~\cite{sommerville2005empirical}. They specify what a system should do, the constraints under which it should operate, and the properties it is expected to satisfy. These descriptive documents are most commonly formalized as \textit{Requirements Specifications}, which establish a shared understanding of the system among all stakeholders involved~\cite{zowghi2005requirements}. Requirements are particularly important because they provide the basis for system design, verification, and validation. Therefore, maintaining accurate and consistent requirements documentation is essential throughout the system lifecycle~\cite{pohl1994three}.

Among the most inevitable realities of any system's lifecycle is change. No system, regardless of how thoroughly it is specified or carefully it is designed, remains static throughout its operational life. Change is driven by different reasons such as evolving regulations, shifting stakeholder needs, discovered system defects, environmental adjustments, technological advancements, or the natural growth of the domain the system serves~\cite{nuseibeh2000requirements}. Such changes extend to the requirements specifications themselves, requiring engineers to revisit, revise, and realign the documentation to accurately reflect the system's new state. Since requirements are interdependent, a change to one requirement can affect other requirements that depend on it~\cite{regnell2008can}.
Maintaining consistency between the evolving system and its requirements documentation demands careful analysis of which requirements are affected and how changes may propagate through their dependencies.  Failure to keep requirements documentation synchronized with actual system changes risks introducing ambiguity, inconsistency, and a growing disconnect between what the system is documented to do and what it actually does. This process is formally known as inter-requirement Change Impact Analysis (CIA), a core practice within the Requirements Engineering (RE) domain~\cite{pohl2016requirements}. 

Performing CIA throughout a system's lifecycle is inherently challenging, particularly in large-scale systems where requirements documentation is extensive. This challenge is particularly pronounced for safety-critical systems due to their additional certification constraints. Under such conditions, the process becomes increasingly time- and resource-intensive, and the likelihood of overlooking affected requirements or misassessing a change's scope also increases. 

To address these challenges, researchers have proposed various techniques to automate the CIA task. Dependency-based approaches rely on dependencies between requirements or other system artifacts to identify impacted requirements and propagate change impact ~\cite{nonsiri2013model,aryani2009change,kchaou2017uml,alkaf2019automated,nejati2016automated,goknil2008change,goknil2011semantics,goknil2014change,rahim2026ontology}. However, structural dependencies alone are generally insufficient to capture the full semantic implications of a change request.
Model-based approaches utilize formal representations of requirements and software artifacts to assess whether they are impacted~\cite{aryani2009change,nejati2016automated,kchaou2017uml,alkaf2019automated}. However, these approaches require requirements and artifacts to conform to specific modeling formalisms, which limits
their applicability across different development contexts. Natural Language Processing (NLP) approaches analyze the linguistic content of requirements and other software artifacts, often leveraging semantic similarity to assess change impact~\cite{arora2015change,arora2015narcia}. However, they often rely on shallow linguistic features that fail to capture deeper semantic relationships, and some of them still require substantial human involvement, limiting scalability and automation.
Furthermore, across these three categories, many existing approaches primarily focus on propagating change impact across requirements and other software artifacts rather than directly identifying impacted requirements in response to a specific change request.

With the recent advent of Large Language Models (LLMs), researchers have explored their application to various software engineering activities. LLMs are advanced 
language models trained on large-scale textual data and capable of understanding and generating human-like text. Their natural language processing capabilities, combined with their ability to be fine-tuned for specialized tasks, make them relevant solutions for requirements engineering activities~\cite{10.1145/3744746,wu2024continuallearninglargelanguage}.
However, the application of LLMs to requirements CIA remains underexplored. 
Only a few approaches have been proposed to address this task~\cite{etezadi2025llm,rusdianto2024innovative}, leveraging prompt engineering and basic RAG techniques.
To the best of our knowledge, no existing work has combined the analytical capabilities of LLMs with the structured representation provided by requirements dependency graphs while also leveraging the technical knowledge embedded in the requirements specification for change impact analysis.

In this paper, we present \textit{DRILS}: Dependency-based Requirement change Impact analysis using LLMs and Requirement-Enrichment RAG (RE-RAG), an automated multi-agent architecture for change impact analysis. \textit{DRILS} includes four steps. The first step is RE-RAG, a novel adaptation of Retrieval-Augmented Generation (RAG), a technique that enhances large language models by retrieving and incorporating relevant external knowledge at inference time. RE-RAG supports context-aware knowledge retrieval by resolving referential ambiguity and enriching requirements with relevant domain information in the form of concise bullet points to reduce noise. The second step clusters requirements by semantic similarity, facilitating analysis of large specifications. In the third step, impact set identification, two generator agents take as input the enriched requirement clusters, the change request, and a Requirement Dependency Graph (RDG) to identify potentially impacted requirements. The RDG represents requirements and their dependencies, providing information that helps identify additional requirements that may be affected by a change. For each predicted impacted requirement, the generator agents provide a rationale explaining the prediction, the location within the requirement where the change should be applied, and a confidence score. This auxiliary information promotes agent
self-reflection and enhances the transparency and credibility of the results. Finally, in the fourth step, a judge agent analyzes the generator agents' outputs to produce the final impact set and a rationale justifying the selection of each requirement. These rationales help requirements analysts interpret the results.

We empirically evaluate \textit{DRILS} on three distinct systems from the automotive domain. The \textit{DRILS} architecture achieves an average $F_1$ score of 80.8\% and a recall of 87\%. This high recall is particularly important for CIA, as all potentially impacted requirements should be identified, while falsely predicted requirements can subsequently be filtered out by requirements analysts reviewing \textit{DRILS} outputs. We also conduct ablation studies to assess the impact of auxiliary information generated by the generator agents, RE-RAG, clustering, and the RDG. The results indicate that each type of auxiliary information contributes to improved impact set identification. Their combined effect is particularly pronounced, increasing average recall from 71\% to 87\% across the three datasets, with an even more substantial improvement on one dataset, where recall increases from 62.5\% to 91.6\%. The results also demonstrate the benefits of RE-RAG, clustering, and the RDG, with the RDG having the most pronounced and consistent effect across datasets. The results show that incorporating the RDG into \textit{DRILS} improves overall performance, raising the $F_1$ score from 77\% to 81\% with a relative gain of 5.2\%.

We compare \textit{DRILS} against four state-of-the-art baseline approaches covering four categories of CIA techniques. The results show that \textit{DRILS} substantially outperforms its closest baseline, ProReFiCIA~\cite{etezadi2025llm}, a recently proposed LLM-based technique, in terms of $F_1$ score. Specifically, the average $F_1$ score increases 
from 55\% to 81\%. Although \textit{DRILS} achieves higher average recall, its main advantage is substantially higher precision, increasing from 41.33\% for ProReFiCIA to 76.3\% for \textit{DRILS}. This higher precision reduces the number of falsely predicted impacted requirements that requirements analysts must review, thereby reducing the manual effort required for CIA.  

The main contributions of this paper are as follows:
\begin{itemize}
    \item We propose RE-RAG, a new approach that leverages Retrieval-Augmented Generation (RAG) to enrich requirements with concise, targeted domain context while minimizing irrelevant noise. 
    \item We incorporate a Requirement Dependency Graph (RDG) into the LLM's reasoning process for CIA, enabling it to leverage structured inter-requirement relationships when assessing the scope and propagation of a change.
    \item We propose a multi-agent architecture that combines multiple generator agents with an LLM-as-a-Judge mechanism, promoting diversity of thought and supporting more robust CIA.
    \item We introduce three newly constructed datasets specifically designed to benchmark and evaluate the CIA task, addressing the lack of publicly available datasets for this task.
\end{itemize}

The remainder of this paper is organized as follows. Section~\ref{sec:approach} presents the \textit{DRILS} architecture and details its steps and components. Section~\ref{sec:evaluation} describes the experimental setup, while Section~\ref{sec:Results} presents and analyzes the evaluation results. Section~\ref{sec:discussion} provides an in-depth discussion of the findings. Section~\ref{sec:threats} discusses the threats to the validity of our study. Section~\ref{sec:related-works} reviews the related work. Finally, Section~\ref{sec:conclusion} concludes the paper.
 \section{Approach}\label{sec:approach}

\textit{DRILS} is an LLM-based multi-agent approach that automatically identifies requirements impacted by a change request. \textit{DRILS} uses several techniques to generate an accurate impact set, including requirement enrichment, clustering, multi-agent generation, and judge-based validation.  
The inputs of \textit{DRILS} are: (1) the list of requirements related to the system under consideration, (2) a Requirement Dependency Graph (RDG) describing the dependencies among these requirements, and (3) a change request for which the impact set should be identified. The output is an impact set specific to the input change request, along with the rationale justifying the selection of each requirement within the impact set. This rationale can be used as additional information to assist requirements engineers in understanding and assessing the resulting impact set. 

\textit{DRILS} has two key features: Requirement-Enrichment RAG (RE-RAG), a carefully engineered RAG technique that provides focused domain context, and an RDG as an additional knowledge source. Requirement dependencies play an important role in multiple requirements engineering activities, particularly change impact analysis \cite{Motger2025,ZHANG201440,GOKNIL2014950,darif2026automating}. Although a change may directly affect a set of requirements, its impact can propagate to other requirements that depend on the initially affected ones. Thus, leveraging an RDG supports accurate change impact analysis by identifying all requirements that may be affected by a proposed change request, reducing the risk of overlooking necessary modifications and helping maintain consistency among requirements.

\begin{figure*}
    \centering
\includegraphics[width=1\textwidth]{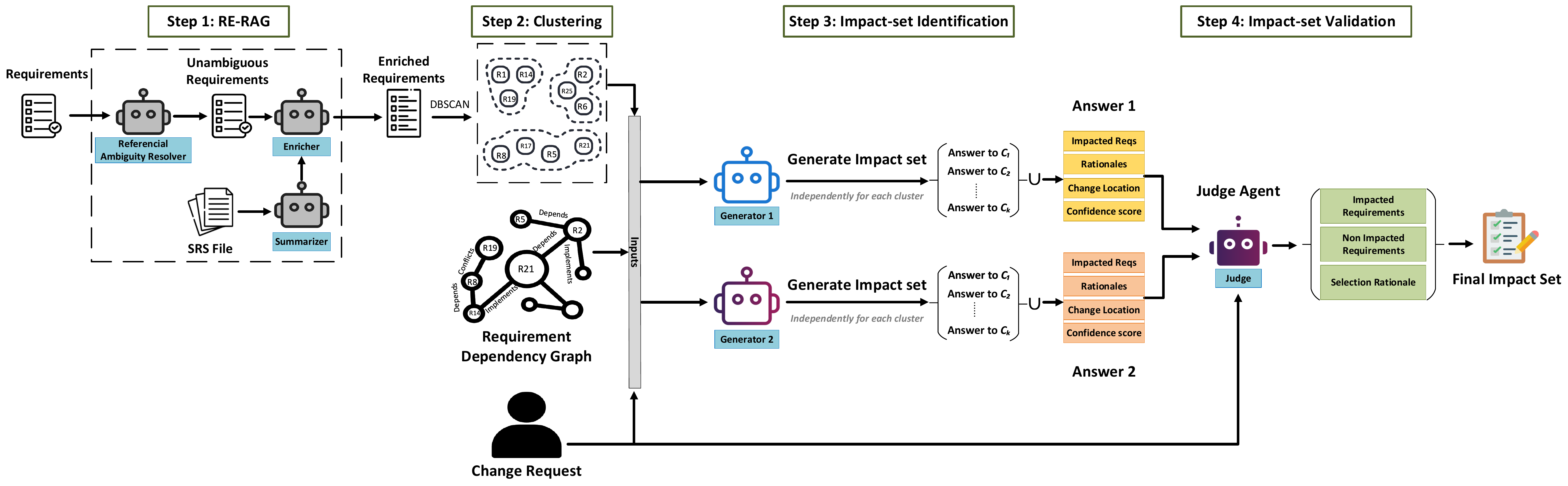}
\caption{Overall architecture of \textit{DRILS}}
    \label{fig:methodology}
\end{figure*}

Figure~\ref{fig:methodology} illustrates \textit{DRILS}'s overall architecture. It includes four steps: (1) RE-RAG, (2) clustering, (3) impact set identification, and (4) impact set validation. The first step involves three agents, which collectively aim to make the requirements as clear and informative as possible. One agent resolves referential ambiguities in the requirements, while another enriches the unambiguous requirements with domain context. The third agent generates summaries of the system's Software Requirements Specification (SRS) document, which the second agent uses to enrich the requirements. SRS documents are structured specifications that define a software system’s functional and non-functional requirements, encompassing system interfaces, constraints, assumptions, dependencies, and use cases, alongside expected behaviors, quality attributes, and operational limitations \cite{iso2018systems}.
This first step produces enriched requirements, which the second step then clusters based on semantic similarity. Next, for each cluster, two generator agents independently identify an impact set, using the RDG and the change request as input. The output impact set includes four elements: (1) the set of impacted requirements, (2) rationales justifying the selection of these requirements, (3) the specific location within the impacted requirements that should be changed, and (4) confidence scores indicating the agent's certainty regarding each identified requirement. This information provides a deeper understanding of the agent's reasoning. 
Finally, in the fourth step, a judge agent evaluates the two agents' outputs and generates the final impact set for the input change request. The final impact set includes the impacted requirements and their corresponding rationales. The rationales aim to facilitate requirement engineers' analysis and assessment of \textit{DRILS} output. The following sections describe each step in detail.

\subsection{Step 1: Requirement-Enrichment RAG (RE-RAG)}

Each requirement in an SRS document describes a specific functionality of the software system. However, extracting a requirement from the SRS without its surrounding context or additional information can make it ambiguous. For example, consider this requirement: \textit{"Allow drivers to provide input during operation, such as speed adjustments or disabling the system altogether, with buttons on the steering wheel."}. Although the requirement describes a specific functionality (i.e., allowing the user to provide an input during operation), it does not explicitly indicate: (1) the entity responsible for enabling the driver to provide input, and (2) the system to which the input applies. Consequently, a reader unfamiliar with the SRS document may be uncertain about which system and entity this requirement involves. This is especially true because, in most SRS documents, the term \textit{''system''} can refer to the overall system being described or its subsystems. This ambiguity can hinder accurate interpretation of the requirement during change impact analysis. A well-known technique that addresses these issues is retrieval-augmented generation.

Retrieval-Augmented Generation (RAG) augments LLMs with relevant domain knowledge extracted from external documents~\cite{10.1145/3744746}. RAG supports task-specific reasoning and mitigates the limitations of relying solely on the model's internal knowledge~\cite{ram2023context}. However, conventional RAG approaches typically retrieve a fixed number of fixed-size chunks based on semantic similarity to the input, without explicitly considering whether the retrieved information is relevant to the specific task~\cite{asai2024self}. Such retrieval can introduce redundant or irrelevant information into the context and potentially hinder the model's ability to identify and use the most relevant information~\cite{asai2024self}. Moreover, increasing the number of retrieved chunks does not necessarily improve LLM performance. LLMs may struggle to use relevant information effectively when it is embedded in longer contexts, degrading performance~\cite {liu2024lost}. These limitations are particularly problematic for change impact analysis, as it requires identifying the technical information needed to determine whether and how a change may affect each requirement. Providing excessive or irrelevant information can substantially increase the complexity of this task. 

Motivated by these limitations, we propose \textbf{Requirement-Enrichment RAG (RE-RAG)}, a RAG-based technique that selectively retrieves relevant information to address specific issues within requirements, such as referential ambiguity and lack of technical details. RE-RAG aims to make each requirement self-explanatory and easier to interpret. To achieve this, RE-RAG includes two sequential stages: (1) Ambiguity Resolution and (2) Contextual Enrichment. To illustrate both stages of RE-RAG, we use the example presented in \autoref{fig:srag-example}. In \textit{DRILS}, we used \texttt{llama3.1-70b} for RE-RAG because of its strong reported performance on natural language processing tasks~\cite{grattafiori2024llama} and its open-source availability, which reduces the cost of running \textit{DRILS}.

\begin{figure}
    \centering
    \includegraphics[width=0.5\textwidth]{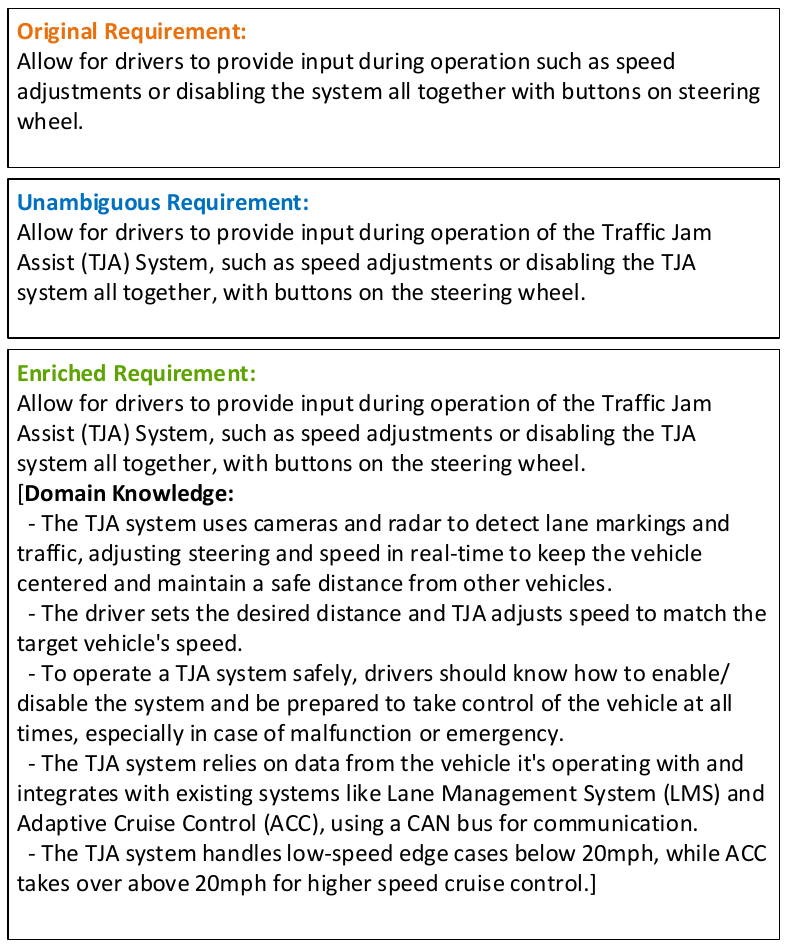}
    \caption{Example of a requirement across different RE-RAG's stages.}
    \label{fig:srag-example}
\end{figure}

\subsubsection{\textbf{Ambiguity Resolution}} This stage aims to resolve all referential ambiguities present in a requirement. Referential ambiguity occurs when a referring expression (e.g., a pronoun or vague noun) does not clearly identify the entity it refers to~\cite{nieuwland2008neurocognition}. 
In well-structured SRS documents, requirements are typically written sequentially, such that references within a requirement can be interpreted unambiguously based on those that precede it. This reflects a natural and intuitive writing practice, consistent with our analysis of multiple public and industrial SRS documents.
Accordingly, we build an \textit{Ambiguity Resolver Agent} that resolves referential ambiguities by providing the target requirement and the five preceding requirements as context. We present the prompt used for this agent in \autoref{fig:AmbiguityResolutionprompt}. Specifically, it instructs the LLM to replace pronouns and vague nouns with explicit referents. The output includes both the unambiguous requirement and the rationale explaining the replacement made. For instance, in \autoref{fig:srag-example}, the \textit{Referential Ambiguity Resolver Agent} resolved two referential ambiguities in the original requirement: which system the operation refers to and which system the user's input affects. The agent identified the “TJA system” as the correct reference in both cases. 

\begin{figure}[ht]
    \centering
    \begin{tcolorbox}[arc=1mm,width=\columnwidth,top=0mm,left=0mm,  right=0mm, bottom=0mm,
                      boxrule=0.3pt,
                   colback=gray!15!white,
                   colframe=black,
                   ]
Below are the previously defined requirements. Use them as context to resolve any referential ambiguity in the current requirement. Replace ambiguous references (e.g., pronouns or vague noun phrases) with their explicit referents whenever they can be inferred from the previous requirements. If no referential ambiguity exists, return the current requirement unchanged. \newline
    \textbf{Previous requirements:} <list> \newline
    \textbf{Current requirement:} <text> \newline
    \textbf{Output Format:} \newline
        - Rationale: <text> \newline
        - Unambiguous requirement: <text>

    \end{tcolorbox}
    \caption{Prompt for the Ambiguity Resolution Agent}
    \label{fig:AmbiguityResolutionprompt}
\end{figure}

\subsubsection{\textbf{Contextual Enrichment}}
This stage enriches each unambiguous requirement produced in the previous stage with relevant domain knowledge. This knowledge is extracted from the SRS document through two steps: summarization and requirement enrichment. In the first step, a \textit{Summarization Agent}  initially extracts all sections from the SRS document. Then, it analyzes each section separately and generates a concise summary capturing its key information. This step uses established text summarization techniques in natural language processing~\cite{kryscinski2019neural} to represent domain information more concisely, reduce redundancy, and remove unnecessary wording. Moreover, the shorter input length lets the model incorporate information from a larger portion of the SRS document while reducing computational costs. This allows the LLM to process domain knowledge more efficiently during enrichment. \autoref{fig:SummarizerPrompt} presents the prompt used by the \textit{Summarization Agent}.

Afterward, during the requirement enrichment step, an \textit{Enrichment Agent} identifies the three summarized sections that are most semantically similar to each unambiguous requirement. These summaries are then incorporated into the prompt as contextual information. Given each requirement and its corresponding selected sections, the \textit{Enrichment Agent} is prompted to generate concise domain knowledge specifically relevant to the requirement. This extracted domain knowledge should clarify ambiguities and fill in missing information in the requirement, facilitating clear and complete understanding. Specifically, the agent is instructed to represent the generated domain knowledge as multiple independent sentences (i.e., bullet points). Based on our observations, this structured representation encourages the LLM to generate concise, self-contained knowledge units that agents can use more readily in the next step than paragraph-based representations. The following presents the prompt used by the \textit{Enrichment Agent}. For example, consider the unambiguous requirement in \autoref{fig:srag-example}. The \textit{Enrichment Agent} generated five bullet points as domain context. The bullet points clarify both how the TJA system operates and how the user (i.e., driver) interacts with it. This information enables a clear and complete understanding of the requirement.

\begin{figure}[ht]
    \centering
    \begin{tcolorbox}[arc=1mm,width=\columnwidth,top=0mm,left=0mm,  right=0mm, bottom=0mm,
                      boxrule=0.3pt,
                   colback=gray!15!white,
                   colframe=black,
                   ]
Summarize the following text by retaining its key information, concepts, and domain knowledge:\newline
    \textbf{Input:} <section>
    \end{tcolorbox}
    \caption{Prompt for the Summarization Agent}
    \label{fig:SummarizerPrompt}
\end{figure}

\begin{figure}[ht]
    \centering
    \begin{tcolorbox}[arc=1mm,width=\columnwidth,top=0mm,left=0mm,  right=0mm, bottom=0mm,
                      boxrule=0.3pt,
                   colback=gray!15!white,
                   colframe=black, 
]
You are provided with multiple sections from a document that are retrieved based on their similarity to a given requirement. For each requirement, identify and extract information from these sections that can enrich the requirement. The extracted information should add domain context, clarify entities or references, resolve ambiguities, explain related functionality or constraints, and provide missing information needed for a more complete understanding of the requirement. Only include information that is directly relevant to the requirement and avoid adding assumptions or unrelated details.\newline
    \textbf{Sections:} <list> \newline
    \textbf{Requirement:} <text> \newline
    \textbf{Output format:} <bullet\_points\_of\_sentences>
    \end{tcolorbox}
    \caption{Prompt for the Enrichment Agent}
    \label{fig:EnrichmentPrompt}
\end{figure}

\subsection{Step 2: Requirements Clustering}
When an analyst provides a change request, the system must consider all requirements in the SRS to determine which requirements may be impacted. Similarly, an LLM needs to examine all requirements in the SRS to identify the potential impact set. However, when an SRS contains a large number of requirements, which is generally the case for safety-critical systems, providing all requirements in a single prompt might not be practical or scalable due to the LLM's context-window limitations. This challenge is even greater in \textit{DRILS} because requirements include additional domain knowledge. This increases the tokens required to represent each requirement and, consequently, reduces the number of requirements that can be processed simultaneously. Furthermore, since the requirements dependency graph is a necessary input to Step 3, randomly dividing requirements into batches may disrupt the graph structure by separating highly dependent requirements across different batches.

To address these challenges, we cluster the unambiguous requirements based on their semantic similarity. During clustering, we exclude domain knowledge because it may introduce information beyond what the requirements define, which could impair clustering by making semantically distinct requirements appear more similar. We perform clustering using DBSCAN (Density-Based Spatial Clustering of Applications with Noise), a density-based clustering algorithm that groups data points by the density of their surrounding neighborhoods without requiring a predefined number of clusters~\cite{ester1996density}. By grouping semantically similar requirements, we aim to form smaller, more coherent clusters while preserving the relationships captured by the requirements dependency graph. This is motivated by the observation that requirements with higher semantic similarity are more likely to exhibit dependencies~\cite{darif2026automating}. Consequently, the resulting clusters are more likely to form well-connected subgraphs, reducing the separation of interdependent requirements across batches and providing the LLM with a more coherent context for impact analysis. It also improves the scalability of requirements change impact analysis. The resulting requirement clusters are provided as input to the third step: \textit{''Impact-Set Identification''}.

\subsection{Step 3: Impact-set Identification}
In the \textit{''Impact-set Identification''} step, two heterogeneous LLMs, \texttt{GPT-4o} and \texttt{Claude-sonnet-5}, are deployed as independent \textit{Generation Agents}. Rather than relying on a single model, this multi-agent setup promotes "Diversity of Thought"~\cite{hegazy2024diversity} by allowing distinct LLMs with different architectures to independently evaluate the input and generate the impact set. We selected these models for their strong reasoning capabilities and widespread adoption in the RE domain~\cite{zadenoori2025large,siqueira2026use}.
By combining models with different reasoning patterns and analytical perspectives, \textit{DRILS} aims to capture a wider range of potential requirement dependencies from the provided contextual knowledge. In this step, both agents take as input: (1) the clusters of enriched requirements, (2) the requirement dependency graph, and (3) the change request. 

Within \textit{DRILS}, the dependency graph is represented as triples, where each triple includes two requirements and the dependency between them, e.g., \texttt{(requirement\_ID1, dependency, requirement\_ID2)}. Triple-based representations are commonly used for graphs, knowledge graphs in particular, and have been shown to be effective as inputs to LLMs \cite{DAI2025113060,yao2025exploringlargelanguagemodels}. This representation isolates individual dependencies and enables agents to consider only the dependencies relevant to the requirements being analyzed. While a dependency graph can include various types of dependencies, \textit{DRILS} focuses on four types that are particularly useful for CIA: \textit{Requires}, \textit{Implements}, \textit{Conflicts}, and \textit{Details}. A \textit{Requires} dependency exists when the fulfillment of a requirement is a prerequisite for the fulfillment of another requirement. An \textit{Implements} dependency exists when a high-level requirement is implemented through a lower-level requirement. A \textit{Conflicts} dependency exists when the fulfillment of one requirement restricts the fulfillment of another requirement. Finally, a \textit{Details} dependency exists when one requirement provides additional details about a shared action described in another requirement. The \textit{Requires}, \textit{Implements}, and \textit{Details} dependencies are directional, while the \textit{Conflicts} dependency is not. We adopted the definitions from the LEREDD framework~\cite{darif2026automating} used in the prompt. 

We selected these dependency types because they are relevant to CIA. Specifically, for the \textit{Requires} dependency, if a required requirement changes, the dependent requirement may also be affected. Furthermore, if either requirement with an \textit{Implements} dependency changes, the other requirement can also be affected by the change: (1) if the higher-level requirement changes, it might render the lower-level requirement inconsistent and thus require modification, and (2) if the lower-level requirement changes, it might no longer implement the higher-level requirement, and thus the latter may be impacted by the change. Similarly, for the \textit{Conflicts} (i.e., Restricts) dependency, modifying one requirement requires verifying the validity of the other, which may therefore be affected by the change. Finally, for the \textit{Details} dependency, if a requirement changes, its detailing requirement might also need to be modified. Incorporating these dependencies helps identify all requirements that may be affected by a change request while reducing the risk of overlooking necessary changes.

Each cluster is processed independently by both agents to generate an impact set. Specifically, each agent runs $k$ times, where $k$ is the number of clusters generated in Step 2. The outputs for each cluster are then concatenated to form the complete impact set. For each requirement included in the impact set, the agents provide: (1) a rationale justifying the selection of the requirement, (2) the location of a suggested change within the requirement, and (3) a confidence score, measured on a 5-point Likert scale, indicating the agent's certainty in its selection. These additional outputs encourage agent self-reflection and reduce the risk of hallucinations. Specifically, instructing agents to generate a rationale can improve their reasoning and promote transparency~\cite{xu2024sayselfteachingllmsexpress,i2023minimizingfactualinconsistencyhallucination}. Moreover, these rationales can help identify issues that may lead to incorrect selections and facilitate their resolution during subsequent \textit{DRILS} refinement iterations. Within the \textit{DRILS} architecture, these rationales are particularly important because they provide the \textit{Judge Agent} in step 4 with additional information about each agent's underlying reasoning, facilitating comparison and identification of the correct answer when agents disagree. Similarly, specifying the location of the change gives the \textit{Judge Agent} additional information about the specific change each agent identifies, enabling it to assess whether the change is logically sound. Finally, instructing the agents to provide confidence scores facilitates the interpretability of their results~\cite{detommaso2024multicalibrationconfidencescoringllms}. These scores can also guide the \textit{Judge Agent} in determining which agent exhibits greater confidence in its answer and is therefore more likely to be correct in cases of conflict between the two agents. Overall, this information enables the \textit{Judge Agent} to assess, compare, and reconcile the agents' individual decisions, ultimately supporting the selection of the final impact set.

\begin{figure*}
    \centering
    \includegraphics[width=1\textwidth]{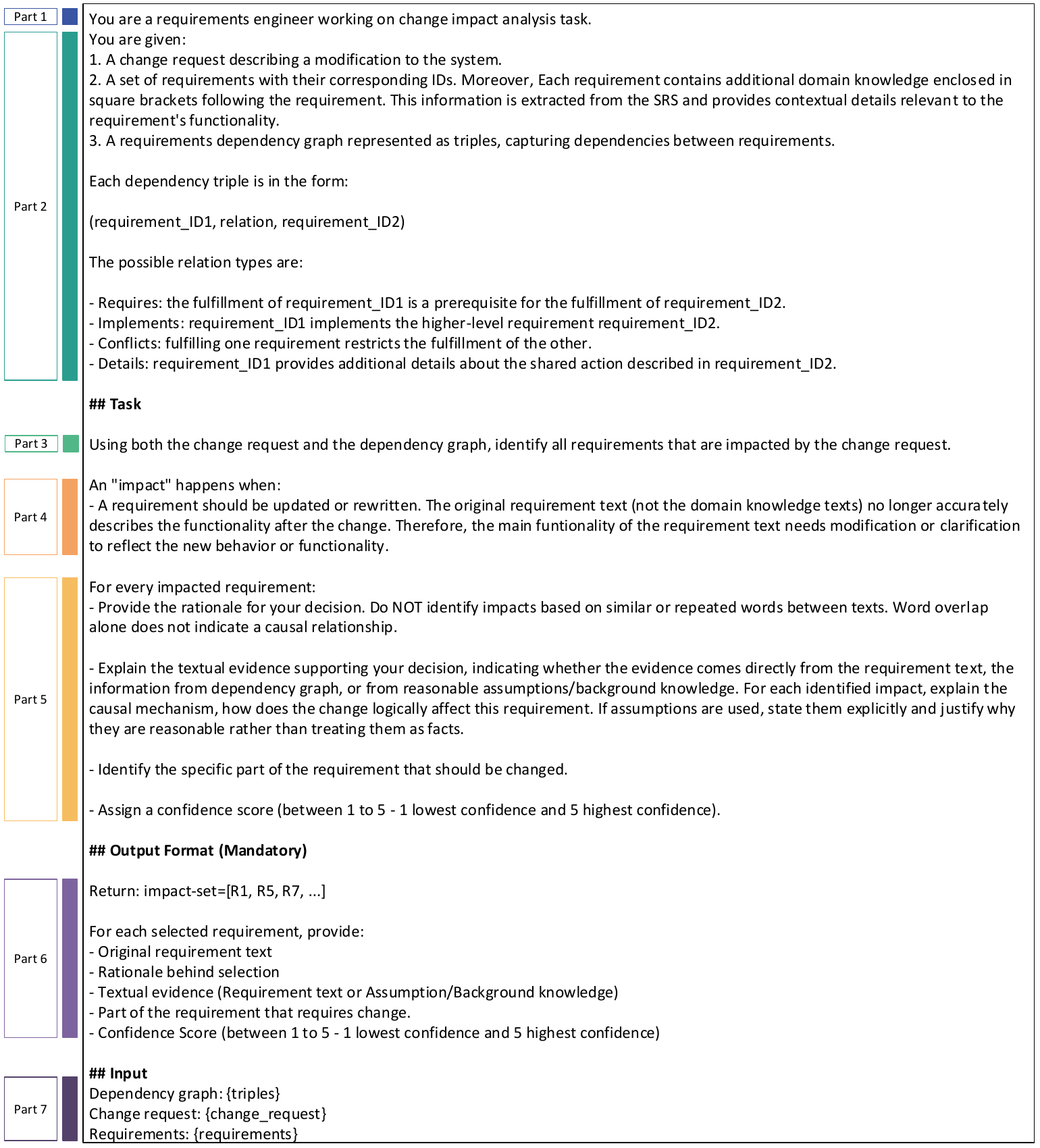}
    \caption{Illustration of the prompt used by the generation agents.}
    \label{fig:main-prompt}
\end{figure*}

To guide both agents in identifying the impact set, we engineered the prompt illustrated in Figure~\ref{fig:main-prompt}. The prompt consists of seven parts. \textit{Part 1} establishes the system persona by assigning the agents the role of a domain-expert requirements engineer tasked with conducting CIA. This role specification gives the model a clear perspective from which to interpret the inputs. \textit{Part 2} introduces and describes the three inputs provided to the agent: the change request, the cluster of enriched requirements, and the requirement dependency graph. It also provides additional information about the dependency graph, including what a requirement dependency graph represents, the different types of dependencies, and the direction of each dependency type. In \textit{Part 3}, we provide the model with a high-level description of the task and instruct it to use the dependency graph to identify the requirements impacted by the given change request. 

\textit{Part 4} defines the term ''impact'' in the context of the CIA task. This part is particularly important because it establishes clear criteria for determining whether a requirement should be considered impacted by the given change request. By explicitly defining impact, we give the model a consistent interpretation of the task and reduce ambiguity in its decision-making. This helps ensure the identified impact set reflects the intended scope of the CIA task rather than the model’s potentially varying interpretation of what constitutes an impact. We define a requirement as impacted when it should be updated or rewritten in response to the change request, as it no longer accurately describes the functionality. 

\textit{Part 5} instructs the agent to provide textual evidence supporting its decisions. To achieve this, we instruct the model to provide: (1) the rationale behind its selection, (2) the textual evidence supporting its decision, (3) the specific location within the impacted requirement that should be changed in response to the given change request, and (4) a confidence score. In particular, regarding textual evidence, the agent must clarify whether its decision is based solely on the requirement text, information from the dependency graph, or its own assumptions and background knowledge. This information not only guides the agent toward more self-aware and transparent responses, but also plays a crucial role, as the Judge Agent in Step 4 of \textit{DRILS} subsequently uses these outputs. \textit{Part 7} describes the format for presenting these outputs. Finally, in \textit{Part 8}, we provide all inputs the agent needs for the task, in the same order they were introduced in \textit{Part 2}.

\begin{figure*}
    \centering
    \includegraphics[width=1\textwidth]{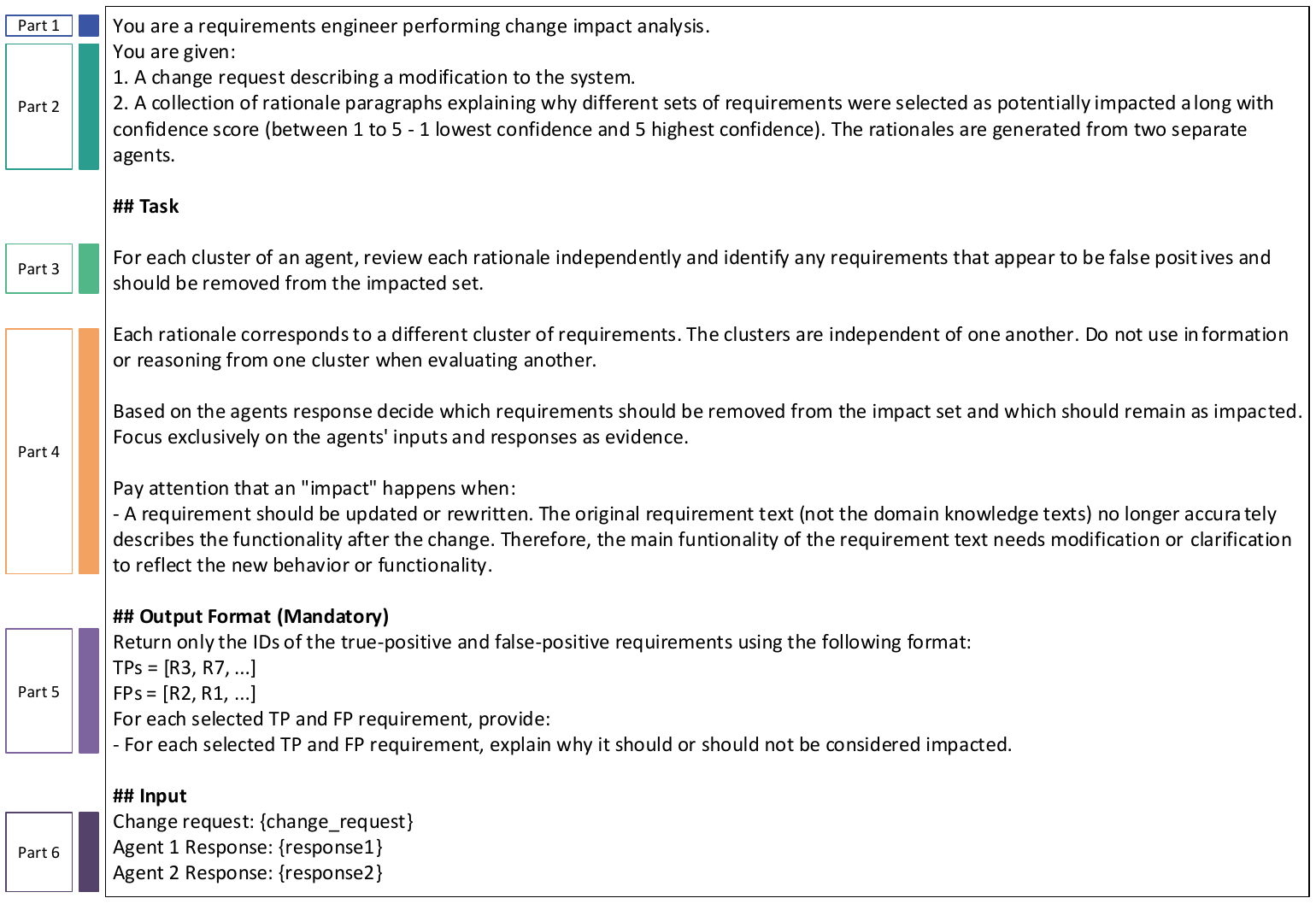}
    \caption{Illustration of the prompt used by the Judge Agent.}
    \label{fig:srag-example}
\end{figure*}

\subsection{Step 4: Impact-set Validation}
Impact-set Validation is the final step of \textit{DRILS}, in which a \textit{Judge Agent} analyzes the outputs of the agents in step 3 and produces the final impact set. LLM-as-a-Judge is a technique in which LLMs are employed as evaluators for complex tasks~\cite{gu2025surveyllmasajudge}. In particular, it is especially relevant for evaluating other LLMs' outputs, as LLM judges have shown substantial agreement with human evaluations, making them a scalable and effective technique for assessing LLM outputs~\cite{gu2025surveyllmasajudge,zheng2023judgingllmasajudgemtbenchchatbot}. To this end, we employ an LLM from a different model family, \texttt{Gemini-3.5-flash}, as our \textit{Judge Agent}. By using an LLM from a different model family than the generation agents, we aim to reduce potential confirmation bias and avoid simply reinforcing the decisions produced during step 3~\cite{zheng2023judging}. \texttt{Gemini-2.5-flash} was selected as the \textit{Judge Agent} due to its strong performance~\cite{siqueira2026use} and large context window, which is essential in \textit{DRILS} where the outputs of two agents are combined and passed as input to the judge.

The \textit{Judge Agent} is responsible for separating the requirements identified by the \textit{Generation Agents} into those it considers true positives (i.e., impacted requirements) and false positives (i.e., non-impacted requirements), with true positives constituting the final impact set. To guide the \textit{Judge Agent} through this task, we engineered the prompt illustrated in Figure~\ref{fig:srag-example}. The prompt includes six parts. \textit{Part 1} establishes the requirements engineer persona for the Judge Agent and guides it to evaluate the candidate impact sets from a change impact analysis perspective. \textit{Part 2} explains the inputs provided to the \textit{Judge Agent}. For each \textit{Generation Agent}, the outputs across all $k$ clusters are concatenated to form the agent's complete response. The change request, along with the concatenated responses from both agents, is then provided as a single combined input to the \textit{Judge Agent}.
\textit{Part 3} provides task-related instructions to the \textit{Judge Agent}. Specifically, it states that some requirements identified by the \textit{Generation Agents} may be false positives and should therefore be removed from the final impact set. \textit{Part 4} provides the \textit{Judge Agent} with a detailed explanation of the input format to prevent incorrect assumptions. It explains that the \textit{Generation Agents} output multiple clusters, each containing a list of impacted requirements, with each requirement accompanied by a rationale, a suggested change location, and a confidence score. Furthermore, the \textit{Judge Agent} bases its decision solely on the information provided by the \textit{Generation Agents} and avoids making assumptions based on information not included in the input. This prevents the \textit{Judge Agent} from making uninformed assumptions during validation. We also define what an impact is to help the Judge Agent produce the final impact set. In \textit{Part 5}, we specify the expected output format, including the true positives (i.e., the final impact set), false positives, and the rationales justifying their selection and exclusion. Finally, \textit{Part 6} provides the Judge Agent with the inputs in the same order as described in Part 2, while separating the responses generated by the two Generation Agents.

 \section{Empirical Evaluation}\label{sec:evaluation}
In this section, we describe the empirical evaluation conducted to assess \textit{DRILS} across different settings and benchmark its performance against established baselines.
To evaluate \textit{DRILS} and assess each component's contribution to overall output quality, we adopt a backward evaluation strategy: we evaluate each component starting from the last one that produces the final output and trace back to the first component that receives the initial input. This approach is motivated by error propagation in multi-component NLP pipelines, where errors introduced by an upstream component can compound through subsequent stages, making it difficult to attribute final-output quality to an individual component when evaluated in a forward manner~\cite{finkel2006solving}. Accordingly, in RQ1, we begin by evaluating the \textit{Judge Agent} (Step 4 in \autoref{fig:methodology}) and examining how variations in its inputs affect final output quality. In RQ2 and RQ3, we then assess the quality of the outputs produced by the \textit{Generator Agents} (Step 3 in \autoref{fig:methodology}) under different variations in the inputs generated by the pre-processing steps (Steps 1 and 2 in \autoref{fig:methodology}), namely RE-RAG, Clustering, and the dependency graph. Finally, in RQ4, we evaluate \textit{DRILS} against a set of recent and widely adopted baseline methods.

\subsection{Research Questions (RQs)}

\textbf{\textit{RQ1. How does the auxiliary information produced by the generator models (i.e., rationale, change location, and confidence score) affect the performance of \textit{DRILS}?}}

In this research question, we investigate how each type of auxiliary information produced by the \textit{Generator Agents} influences the \textit{Judge Agent}'s final decision. Specifically, we investigate the effects of the rationale, change location, and confidence score. This RQ is divided into four sub-RQs, each assessing the individual effect of a single auxiliary information, as well as their combined effect.

\begin{itemize}
    \item \textbf{\textit{RQ1.1 How does the rationale affect the performance of \textit{DRILS}?}}

    \item \textbf{\textit{RQ1.2. How does the change location affect the performance of \textit{DRILS}?}}

   \item \textbf{\textit{RQ1.3. How does the confidence score affect the performance of \textit{DRILS}?}}

   \item \textbf{\textit{RQ1.4. How does removing all auxiliary information affect the performance of \textit{DRILS}?}}
 
\end{itemize}

\textbf{\textit{RQ2. How do the pre-processing stages, i.e., RE-RAG and clustering, affect performance of \textit{DRILS}?}}

In this research question, we investigate the extent to which the pre-processing stages of RE-RAG and Clustering contribute to the quality of the outputs generated by the overall \textit{DRILS} architecture. This RQ is divided into two sub-RQs, each comparing the adopted strategy with alternative approaches: (1) RE-RAG with traditional chunk-based RAG and No RAG, and (2) Clustering with listing all requirements and a fixed batching strategy.

\begin{itemize}
    \item \textbf{\textit{RQ2.1 How does RE-RAG affect the performance of \textit{DRILS} compared with traditional chunk-based RAG and No RAG?}}

    \item \textbf{\textit{RQ2.2. How does Clustering affect the performance of \textit{DRILS} compared with full requirements listing and fixed batching?}}

\end{itemize}

\textbf{\textit{RQ3. How does providing an RDG as input affect the performance of \textit{DRILS}?}}

In this research question, we investigate the extent to which providing the requirements dependency graph as input to the generator agents contributes to the quality of DRILLS outputs.

\textbf{\textit{RQ4. How does \textit{DRILS} perform compared with baseline approaches?}}

In this research question, we compare the performance of \textit{DRILS} against three widely adopted and recently proposed techniques for the CIA task, covering three types of approaches: (1) Latent Semantic Indexing (LSI) as an information retrieval approach, (2) SBERT and LLM-Vec as pre-trained embedding approaches, and (3) ProReFiCIA as an LLM-based approach.

\begin{figure*}
    \centering
\includegraphics[width=0.9\linewidth]{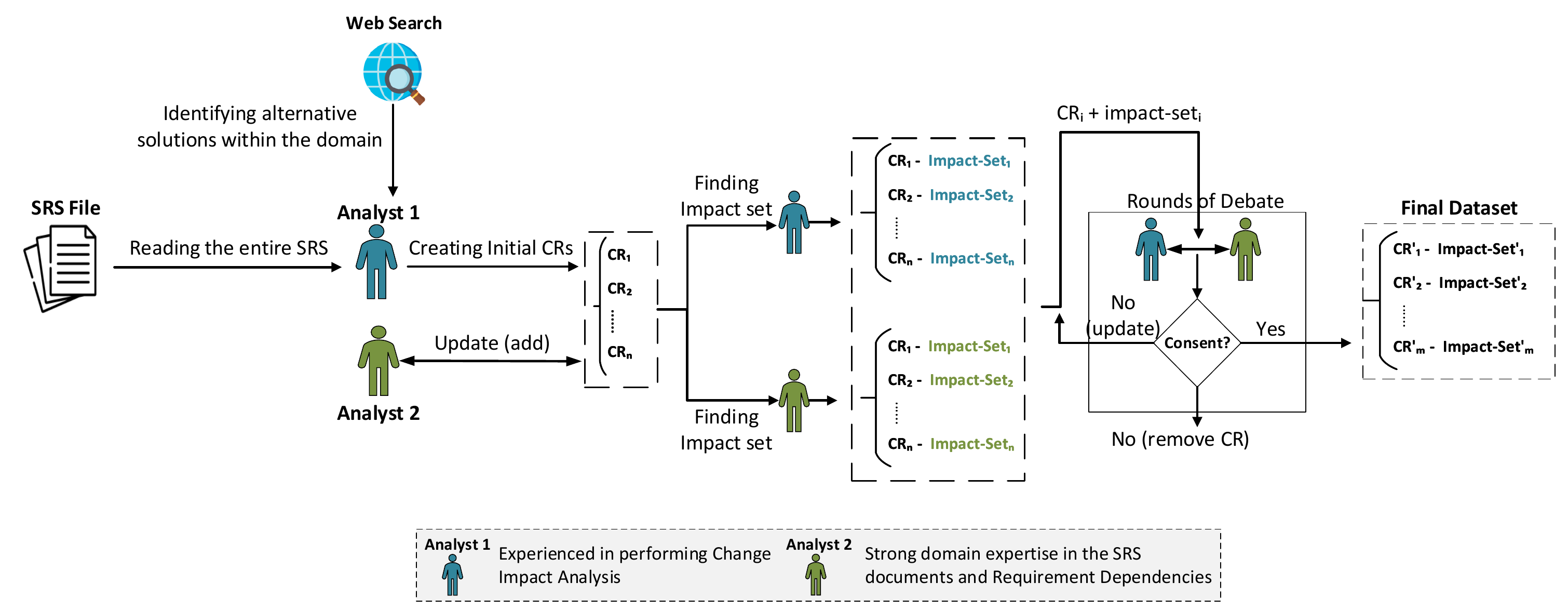}
    \caption{Dataset Construction Process.
}
    \label{fig:dataset-creation}
\end{figure*}

\subsection{Datasets}\label{sec:dataset}
For the empirical evaluation, we use a set of publicly available Software Requirements Specification (SRS) documents that were developed as part of the Requirements Engineering course at  Michigan State University\footnote{{\url{https://www.cse.msu.edu/~cse435/\#cinfo}}}. These documents were developed through an industry-informed requirements engineering curriculum and benefited from direct involvement of industrial partners throughout their development and validation. Repeated cycles of review and refinement with industrial partners produced specifications that accurately capture real industrial practices, making them suitable for the present empirical evaluation. 

From the available course documents, we selected three SRS documents covering distinct automotive safety-critical systems: (1) Automated Parking Assist (APA), a system that supports automated vehicle parking maneuvers, (2) Traffic Jam Assist (TJA), a system that maintains a safe distance from nearby vehicles in heavy traffic by detecting surrounding vehicles and lane markings, and (3) Adaptive Driving Beam (ADB), a system that automatically adjusts vehicle headlights based on driving conditions. We select these documents because prior work~\cite{darif2026automating} manually extracted and validated dependencies among the requirements in these documents. These dependency annotations are particularly valuable given the scarcity of publicly available requirements datasets that are annotated for dependencies. The dependencies identified in this work~\cite{darif2026automating} were provided as input to the \textit{Generator Agents} in the third step of \textit{DRILS}. Note that each system is represented by a separate dependency graph.

For each SRS document, we manually created a set of change rationales through a systematic process. Figure~\ref{fig:dataset-creation} illustrates this process. Two analysts, both authors of this paper, participated in constructing the dataset. In the first stage, \textit{Analyst 1}, who has experience in change impact analysis, thoroughly reviewed each SRS document to develop a comprehensive understanding of the corresponding automotive system. External resources, including targeted web searches, were also consulted to identify practically relevant and feasible system changes. For example, the SRS of the APA system specifies that: \textit{“the communication between subsystems is via the Controller Area Network (CAN) bus”}. To develop a realistic change rationale for this statement, \textit{Analyst 1} consulted automotive technical resources to: (1) identify alternative communication technologies that could potentially replace the CAN bus, and (2) understand the potential system-level implications of such change.

Following this procedure, we formulated an initial set of change rationales and then had \textit{Analyst 2}, who has previously analyzed these documents and thus acquired relevant domain expertise, review them. \textit{Analyst 2} further refined and expanded the change rationales to improve their practical relevance. Finally, the two analysts independently identified the potential impact set for each change rationale. Afterwards, the final impact set for each change rationale was established through multiple rounds of discussion and deliberation, with both analysts reviewing their selections of impacted requirements until reaching a consensus. The annotated datasets are publicly available at \cite{annex}.

Table~\ref{tab:datasets} presents the characteristics of the datasets used in the empirical evaluation: D1 (APA), D2 (TJA), and D3 (ADB). For each dataset, the table reports: (1) the number of requirements, (2) the number of change rationales, (3) the Mean Impact Density (\textit{MID}), and (4) the percentage of requirements impacted by at least one change rationale (\textit{P}). The Mean Impact Density (MID) measures the average proportion of requirements impacted by a change rationale within each dataset. Specifically, for each change rationale, the impact density is calculated as the ratio of the number of impacted requirements to the total number of requirements in the dataset. \textit{MID} is then obtained by averaging these values across all change rationales. 

As Table~\ref{tab:datasets} shows, \textit{MID} is below 7\% across all datasets, indicating that individual change rationales impact only a small fraction of requirements. This demonstrates the sparsity of requirements impacted by change requests. Therefore, given a potentially large set of requirements, an analyst must identify a relatively small subset that is actually affected by the proposed change. This sparsity makes the CIA task challenging and highlights the importance of approaches that can automatically and effectively distinguish impacted requirements from a large number of non-impacted candidates. They can substantially reduce the effort and time required to perform the task, thereby providing significant practical value.

We also report \textit{P}, which calculates the percentage of requirements that are impacted by at least one change rationale across the entire dataset. Specifically, \textit{P} measures the proportion of requirements that appear in the impact set of one or more change rationales, thereby indicating the overall coverage of the requirements by the collected change rationales. As shown in Table~\ref{tab:datasets}, the \textit{P} values are 30\%, 52\%, and 51\% for the APA, TJA, and ADB datasets, respectively. This variation indicates differences in the breadth of change coverage across the datasets. Those \textit{P} values indicate that a substantial portion of requirements are not impacted by any change rationale, further highlighting the need for approaches that can automate the CIA task.

\begin{table}[h]
\caption{Datasets Information.}
\label{tab:datasets}
\begin{tabularx}{\columnwidth}{
    @{\hskip 0.5em}
    >{\raggedright\arraybackslash}m{0.1\columnwidth}
@{\hskip 1em}
    *{4}{>{\centering\arraybackslash}X}
    @{\hskip 0.5em}
}
    \toprule
    \textbf{Dataset} & 
\textbf{\#Req} & \textbf{\#C} & \textbf{MID} & \textbf{P} \\
    \midrule
    \textbf{D1} &
67 & 10 & 4\% & 30\% \\
    \textbf{D2} &
31 & 9 & 7\% & 52\% \\
    \textbf{D3} &
40 & 12 & 6\% & 50\% \\
    \bottomrule
\end{tabularx}

\begin{tablenotes}
    \item \textit{\textbf{Req:} Number of requirements,
    \textbf{C:} Number of change rationales,
    \textbf{MID:} Mean Impact Density,
    \textbf{P:} Percentage of the requirements impacted by at least one change rationale.}
\end{tablenotes}
\end{table}

\subsection{Baselines}\label{sec:baselines}

In order to provide a comprehensive assessment of \textit{DRILS} effectiveness, we evaluate it against multiple baseline approaches. To ensure a fair and relevant comparison, we establish three criteria to guide the selection of appropriate baselines. First, the baselines should represent different types of techniques, providing coverage of a broad range of widely used and state-of-the-art (SOTA) approaches. Second, the baselines should not require human-in-the-loop intervention, as such involvement may not be consistently feasible in practical settings. Moreover, given the importance of automation for CIA, we compare \textit{DRILS} against fully automated techniques. Third, the baselines should not require large amounts of training data, due to the sparsity and limited size of datasets in real-world practice.

Given these criteria, we select three baselines: (1) Latent Semantic Indexing (LSI), an information retrieval–based approach that constructs a domain-specific semantic space based on word occurrence and distribution, (2) SBERT and LLM-Vec, two general-purpose, pre-trained embedding models that generate semantic representations of the input change rationale and requirements, and (3) ProReFiCIA, an LLM-based technique specifically designed for the CIA task. These approaches span different methodological paradigms and do not require human intervention or task-specific training data. The following sections describe these approaches in detail. 

\subsubsection{\textbf{Latent Semantic Indexing}} Latent Semantic Indexing (LSI) is an information retrieval technique that identifies latent semantic relationships between documents and terms (i.e., words) by analyzing the patterns of word occurrence~\cite{deerwester1990indexing}. LSI applies Singular Value Decomposition (SVD) to a term–document matrix to project the original high-dimensional representation into a lower-dimensional semantic space. Documents can then be compared based on their proximity in this latent space, allowing LSI to capture relationships between terms even when they do not share the exact same words. 
To apply LSI for our evaluation, we extract text from each dataset's SRS and use it to train the LSI model, allowing it to learn domain-specific word relationships and build a semantic space. We then use this semantic space to represent both the change rationale and the requirements. Finally, we identify the requirements with the highest similarity scores to the change rationale as the impacted requirements.

\subsubsection{\textbf{Pre-trained Embeddings}} Embeddings represent textual inputs as fixed-dimensional vectors in a continuous semantic space, where vector proximity measures semantic similarity rather than relying solely on lexical overlap between strings \cite{reimers2019sentencebertsentenceembeddingsusing}. These embeddings can be obtained from general-purpose language models or large language models trained on large-scale textual datasets. Since our baseline criteria exclude models that require a training set for fine-tuning, we select two pre-trained embedding models as baselines: (1) SBERT, which is based on a language model, and (2) LLM-Vec, which is based on a large language model. 
Both models are used to assess the effectiveness of general-purpose pre-trained embeddings. Considering both models supports a more comprehensive comparison across different model families. To apply these models for our evaluation, we compute the proximity between the change request vector and the vectors of all requirements, and select the requirements with the highest similarity scores as the impacted requirements.

For both LSI and pre-trained embedding models, the impact set is identified based on similarity scores. Therefore, a threshold strategy is required to determine the cutoff point above which requirements are considered impacted. To enable a comprehensive evaluation, we consider two threshold strategies, $T_1$ and $T_2$, which were selected based on their reported performance in previous studies \cite{eteclassifier,etezadi2025llm,sauvola2000adaptive}. For $T_1$, requirements are ranked in descending order based on their cosine similarity scores with the change rationale. The cutoff point is then determined by identifying the largest drop between two consecutive similarity scores in the ranked list. The higher similarity score associated with this drop is selected as the threshold. All requirements with similarity scores above this threshold are selected as the final impact set. For $T_2$, the threshold is based on the mean and standard deviation of all similarity values produced. Specifically, we calculate the mean similarity score across all requirements, along with the standard deviation. A requirement is considered impacted if its similarity score is above the threshold of \texttt{$\text{mean} + \text{standard deviation}$}. This strategy adapts the threshold to the score distribution associated with each change rationale rather than relying on a fixed threshold. Both threshold strategies are applied to LSI, SBERT, and LLM-Vec.

\subsubsection{\textbf{ProReFiCIA}} ProReFiCIA is an LLM-based approach that identifies change-impacted requirements~\cite{etezadi2025llm}. ProReFiCIA first explores different LLMs and prompt configurations to identify the most effective model–prompt combination, which is then applied to the complete set of requirements to generate an initial impact set. We then refine the result by applying the previously identified optimal configuration to the remaining requirements not identified in the initial pass, mitigating the lost-in-the-middle effect~\cite{liu2024lost} and recovering potentially missed impacted requirements. The candidate impact sets are subsequently combined and passed to a filtering stage, where an LLM ranks the candidates according to the LLM's confidence in their impact. In this stage, a Natural Language Inference (NLI)~\cite{jm3} model also filters the ranked requirements. The NLI model is fine-tuned using change requests as premises and candidate requirements along with their justifications as hypotheses, enabling it to filter out irrelevant candidates. Combining the ranking stage with NLI-based filtering in order to improve precision. We use ProReFiCIA as one of our baselines because it is a recently proposed, well-performing LLM-based approach for the CIA task. It focuses on improving recall, given the importance of identifying impacted requirements. This focus makes ProReFiCIA relevant to our evaluation.  To apply ProReFiCIA, we use the same model–prompt combination introduced in the original paper~\cite{etezadi2025llm}. To fine-tune the NLI model, we employ a leave-one-out strategy, where the model is fine-tuned on two datasets and evaluated on the remaining dataset.

\subsection{Implementation}\label{sec:implementation}

This section describes the implementation details of the \textit{DRILS} components as well as the baseline approaches. We have made the replication package publicly available~\cite{annex}.

\sectopic{RE-RAG.} The agent employed for RE-RAG was \texttt{LLaMa 3.1 70B}. The model was deployed locally using the \texttt{Ollama v(0.6.1)} library across two NVIDIA L40S GPUs, each equipped with 48 GB of RAM. For such smaller LLMs, temperatures that are superior to zero introduce a degree of variability that can be beneficial for text retrieval and reasoning tasks~\cite{li2025exploring}. We set a temperature of 0.4 to maintain moderate variability.

\sectopic{Generator Agents.} \texttt{Claude Sonnet} and \texttt{GPT-4o} were employed as the Generator Agents in \textit{DRILS}, accessed via API calls to Anthropic and OpenAI, respectively. Both models were configured with a \texttt{max\_output\_tokens} limit of 4,000. For \texttt{GPT-4o}, we set a temperature of 0 as it provides more deterministic outputs, which are suitable for its large-scale capabilities as well as the task at hand which requires strong instruction-following and stable outputs~\cite{li2025exploring}.
The \texttt{Claude} API does not support temperature configuration. All API interactions were managed through the \texttt{LangChain v(1.3.4)} library.

\sectopic{Judge Agents.} \texttt{gemini-3.5-flash} was used as the Judge Agent, accessed via API calls to Google. The model was configured with a temperature of 0  to reduce output variability and ensure consistency, and a \texttt{max\_output\_tokens} limit of 4,000. All API interactions were managed through the \texttt{LangChain v(1.3.4)} library.

\sectopic{LSI.} The LSI model was implemented using the \texttt{scikit-learn v(1.9.0)} library. The model was trained on all sections of the SRS documents. We first preprocessed the input texts through normalization and lemmatization. A term-document matrix was then constructed using TF-IDF weighting. The similarity between a query requirement and candidate requirements was measured using cosine similarity.

\sectopic{SBERT.} The SBERT baseline was implemented using the  \texttt{SentenceTransformer v(5.5.1)} library. Specifically, we used the \texttt{\seqsplit{all-MiniLM-L6-v2}}, a pretrained BERT-based sentence encoder, to generate the embeddings for requirements and change requests. We then computed the cosine similarity between the change request and all requirement pairs. We selected this model for its well-demonstrated balance of computational efficiency and effectiveness in capturing semantic similarity~\cite{reimers2019sentence}.

\sectopic{LLM-Vec.} The LLM-Vec baseline was implemented using the  \texttt{\seqsplit{SentenceTransformers v(5.5.1)}} library. The \texttt{\seqsplit{infgrad/Jasper-Token-Compression-600M}} embedding model was selected because it achieved the highest average score across tasks on the Massive Textual Embedding Benchmark (MTEB) leaderboard~\footnote{\url{https://huggingface.co/spaces/mteb/leaderboard}}. The model was used to encode both change requests and requirements into dense vector embeddings, with pairwise similarity computed using the cosine similarity metric.

\sectopic{ProReFiCIA.} ProReFiCIA was implemented in accordance with the settings reported in the original paper~\cite{etezadi2025llm}.

\subsection{Metrics}\label{sec:metrics}

For the empirical evaluation, we employ a diverse set of metrics, covering three categories: (1) performance metrics, which evaluate how well a certain configuration or approach performs on the CIA task, (2) operational metrics, which assess the practical effectiveness and cost of an approach, and (3) a statistical test, which evaluates whether the differences in performance between the evaluated approaches are statistically significant. As performance metrics, we compute \textit{Recall}, \textit{Precision}, and \textit{$F_1$ score}. As operational metrics, we compute  \textit{Effectiveness} and the \textit{Cost} for the analyst to validate the predicted impact set. Finally, we use \textit{Fisher's Exact test}~\cite{fisher1922interpretation} to test statistical significance. The description of these metrics is provided in the following sections.

\subsubsection{\textbf{Performance Metrics}}
\noindent
\sectopic{Recall (R).} Recall measures the proportion of actual positive instances (i.e., impacted requirements) that a model correctly identifies, reflecting its ability to avoid missing relevant cases. The recall is formally expressed in \autoref{equ:recall}, where true positives (TP) represent correctly identified impacted requirements and false negatives (FN) represent impacted requirements that are incorrectly identified as non-impacted.
\begin{equation}\label{equ:recall}
    Recall = \frac{TP}{TP+FN}
\end{equation}

\sectopic{Precision (P).} Precision measures the proportion of predicted positive instances (i.e., impacted requirements) that are actually correct, reflecting the model's ability to avoid incorrectly identifying non-impacted requirements as impacted. Precision is formally expressed in \autoref{equ:precision}, where true positives (TP) are correctly identified impacted requirements and false positives (FP) are incorrectly flagged impacted requirements.
\begin{equation}\label{equ:precision}
    Precision = \frac{TP}{TP+FP}
\end{equation}

\sectopic{$F_1$ score (F1).} $F_1$ score is the harmonic mean of precision and recall, providing a single balanced metric that accounts for both false positives (FP) and false negatives (FN). The $F_1$ score is formally expressed in \autoref{equ:f1}.
\begin{equation}\label{equ:f1}
    F1 = \frac{2 \times Recall \times Precision}{Recall+Precision}
\end{equation}

\subsubsection{\textbf{Operational Metrics}}
\noindent
\sectopic{Effectiveness (Eff).} Effectiveness assesses how well the retrieved impact set captures all impacted requirements. It is measured using macro-averaged recall, which averages recall across all change requests, ensuring each change request contributes equally to the final score regardless of how many impacted requirements it contains~\cite{etezadi2025llm}. This metric represents, on average, the proportion of impacted requirements successfully identified within the impact set. Effectiveness is formally expressed in \autoref{equ:eff}, where $N_c$ represents the total number of change requests in the dataset, $TP_{c_i}$ denotes the number of true positives in the impact set for a change request $c_i$, and $FN_{c_i}$ denotes the number of false negatives, i.e., impacted requirements that were not included in the impact set for change request $c_i$.
\begin{equation}\label{equ:eff}
    Effectiveness = \frac{1}{N_c} \sum_{i=1}^{N_c} \frac{TP_{c_i}}{TP_{c_i}+FN_{c_i}}
\end{equation}

\sectopic{Cost.} Cost is a surrogate measure of the effort required for the requirements analyst to validate the identified impact set, quantified as the proportion of requirements that must be reviewed~\cite{etezadi2025llm}. Cost is computed by calculating the macro-averaged proportion of retrieved requirements relative to the total number of requirements in the dataset, averaged across all change requests. A lower cost indicates that the model retrieves a more focused and compact impact set, thereby reducing the manual validation burden on the requirements analyst. The cost is formally expressed in \autoref{equ:cost}, where $N_c$ represents the total number of change requests in the dataset, $TP_{c_i}$ and $FP_{c_i}$ represent the number of true positives and false positives retrieved in the impact set for a change request $c_i$, respectively, and $N_{req}$ denotes the total number of requirements in the dataset.
\begin{equation}\label{equ:cost}
    Cost = \frac{1}{N_c} \sum_{i=1}^{N_c} \frac{TP_{c_i}+FP_{c_i}}{N_{req}}
\end{equation}

\subsubsection{\textbf{Statistical Analysis}}
\noindent
\sectopic{Fisher's Exact test.} Fisher's exact test is a statistical test used to determine whether there is a statistically significant association between two categorical variables~\cite{fisher1922interpretation}. This test is primarily designed for datasets with small sample sizes, sparse data, or when more than twenty percent of the table cells have expected counts below five. Moreover, testing for an association between two categorical variables using Fisher’s Exact Test is mathematically equivalent to testing whether the proportion of an outcome differs between two groups. Therefore, we use this test to assess whether the differences in recall and precision between two approaches are statistically significant. Specifically, it determines whether the observed differences in precision and recall are associated with differences in the TP versus FP (for precision) and TP versus FN (for recall) proportions, respectively, under the null hypothesis that the two approaches yield identical proportions.

\section{Empirical Results}\label{sec:Results}

\subsection{RQ1: Effect of Auxiliary Information Produced by the Generator Models on \textit{DRILS} Performance}\label{sec:RQ1}
This RQ investigates the impact of the auxiliary information that the \textit{Generator Agents} provide as input to the \textit{Judge Agent} and the extent to which it affects its final impact set selection. The auxiliary information refers to the supplementary outputs generated by the \textit{Generator Agents}, beyond their prediction of whether a requirement is impacted by the change request. This information was explicitly requested in the main prompt, as illustrated in \autoref{fig:main-prompt}, and comprises three elements: (1) \textit{Rationale (R)}, which provides textual evidence explaining the reason behind the agent's selection of a given requirement, (2) \textit{Change Location (CL)}, which specifies the specific part of the requirement that should be modified in response to the given change request, and (3) \textit{Confidence Score (CS)}, a value ranging from 1 to 5 that reflects the agent's confidence in its prediction. 

To examine the effect of each of these elements on the \textit{Judge Agent}'s decision-making, we define five experimental settings in which different auxiliary information elements are excluded from the \textit{Judge Agent}'s input: (1) \textit{DRILS-w/o-R}, where only the rationale is removed, (2) \textit{DRILS-w/o-CL}, where only the change location is removed, (3) \textit{DRILS-w/o-CS}, where only the confidence score is removed, (4) \textit{DRILS-w/o-(R-CL-CS)}, where the rationale, change location, and confidence score are removed, and (5) \textit{DRILS}, where none of this information is removed. To implement these variations, for each dataset, we first run the first three steps of \textit{DRILS} and store the \textit{Generator Agents}' outputs. We then iteratively remove the auxiliary information corresponding to each experimental setting from these stored outputs before passing them to the \textit{Judge Agent}. Therefore, we re-execute only the \textit{Judge Agent} across different input variations. This ensures that the only difference between the experimental settings is the presence or absence of the auxiliary information in question. \autoref{tab:judge-information} presents results for the five experimental settings across recall, precision, $F_1$ score, effectiveness, and cost. 

\begin{table}
\centering
\caption{Experimental results on the effect of auxiliary information on the Judge Agent}
\label{tab:judge-information}
\begin{tabularx}{1\textwidth}{
    @{\hskip 0.1em}
    >{\raggedright\arraybackslash}m{0.05\textwidth}
    >{\raggedright\arraybackslash}m{0.05\textwidth}
    @{\hskip 0.5em}
    *{5}{>{\centering\arraybackslash}X}
    @{\hskip 0.1em}
}
    \toprule
    \textbf{Dataset} & \textbf{Metric} & \textbf{DRILS-w/o-R} & \textbf{DRILS-w/o-CL} & \textbf{DRILS-w/o-CS} & \textbf{DRILS-w/o-(R-CL-CS)} & \textbf{DRILS} \\

    \midrule
    \multirow{5}{*}{\textbf{D1}}
    & \textbf{R}    & \textbf{91.6 $\pm$ 0.00} & $84.7 \pm 0.03$ & \textbf{91.6 $\pm$ 0.00} & $62.5 \pm 0.00$ &  \textbf{91.6 $\pm$ 0.00} \\
    & \textbf{P}    & $61.6 \pm 0.03$ & \textbf{75.3 $\pm$ 0.03} & $69.1 \pm 0.01$ & $70.4 \pm 0.03$ & $64.9 \pm 0.05$ \\
    & \textbf{F1}   & $73.7 \pm 0.01$ & \textbf{79.7 $\pm$ 0.03} & $78.9 \pm 0.01$ & $66.2 \pm 0.01$ & $75.9 \pm 0.03$ \\
    & \textbf{Eff}  & \textbf{94.9 $\pm$ 0.00} & $92.2 \pm 0.01$ & \textbf{94.9 $\pm$ 0.00} & $76.7 \pm 0.00$ & \textbf{94.9 $\pm$ 0.00} \\
    & \textbf{Cost} & $05.6 \pm 0.06$ & $04.1 \pm 0.00$ & $04.7 \pm 0.02$ & \textbf{03.2 $\pm$ 0.03} & $05.1 \pm 0.05$ \\

    \midrule
    \multirow{5}{*}{\textbf{D2}}
    & \textbf{R}    & \textbf{85.0 $\pm$ 0.00} & \textbf{85.0 $\pm$ 0.00} & \textbf{85.0 $\pm$ 0.00} & $78.3 \pm 0.04$ & $83.3 \pm 0.03$ \\
    & \textbf{P}    & $58.6 \pm 0.00$ & $83.7 \pm 0.06$ & $72.9 \pm 0.05$ & $61.0 \pm 0.01$ & \textbf{86.2 $\pm$ 0.03} \\
    & \textbf{F1}   & $69.4 \pm 0.00$ & $84.3 \pm 0.03$ & $78.5 \pm 0.03$ & $68.6 \pm 0.02$ & \textbf{84.8 $\pm$ 0.03} \\
    & \textbf{Eff}  & \textbf{83.3 $\pm$ 0.00} & \textbf{83.3 $\pm$ 0.00} & \textbf{83.3 $\pm$ 0.00} & $79.6 \pm 0.02$ & $82.5 \pm 0.01$ \\
    & \textbf{Cost} & $10.4 \pm 0.00$ & $07.3 \pm 0.05$ & $08.4 \pm 0.05$ & $09.2 \pm 0.02$ & \textbf{06.9 $\pm$ 0.03} \\

    \midrule
    \multirow{5}{*}{\textbf{D3}}
    & \textbf{R}    & $79.0 \pm 0.02$ & $82.7 \pm 0.03$ & $79.0 \pm 0.03$ & $74.1 \pm 0.00$ & \textbf{86.4 $\pm$ 0.02} \\
    & \textbf{P}    & $70.4 \pm 0.06$ & \textbf{88.6 $\pm$ 0.08} & $85.4 \pm 0.02$ & $77.9 \pm 0.02$ & $77.8 \pm 0.04$ \\
    & \textbf{F1}   & $74.4 \pm 0.04$ & \textbf{85.4 $\pm$ 0.03} & $82.0 \pm 0.01$ & $75.9 \pm 0.01$ & $81.9 \pm 0.03$ \\
    & \textbf{Eff}  & $82.2 \pm 0.01$ & $87.5 \pm 0.03$ & $82.4 \pm 0.01$ & $83.1 \pm 0.00$ & \textbf{91.7 $\pm$ 0.01} \\
    & \textbf{Cost} & $06.3 \pm 0.05$ & $05.3 \pm 0.09$ & \textbf{05.2 $\pm$ 0.04} & $05.3 \pm 0.02$ & $06.2 \pm 0.03$ \\
    
    \bottomrule
\end{tabularx}

\end{table}

Note that these experiments specifically target the sensitivity of the \textit{Judge Agent} to each element of input auxiliary information and do not examine how removing such information from the \textit{Generator Agents}' prompts affects their impact set identification. This distinction matters because prior studies show that, even when the input and prompt remain unchanged, variations in output format or the inclusion of additional information within the output structure can affect LLM performance~\cite{xu2024sayself}. Furthermore, for this RQ and all remaining RQs, we ran each configuration three times to ensure experimental robustness and reliable results. The reported results represent the mean across the three runs, along with the corresponding standard deviation.

\subsubsection{\textbf{RQ1.1. Effect of Rationale on the Judge Agent}}
As \autoref{tab:judge-information} shows, when comparing \textit{DRILS} with \textit{DRILS-w/o-R}, the average $F_1$ score across the three datasets decreases from 80.9\% to 72.5\%, a 10.4\% relative decrease that indicates substantial performance degradation when the rationale is removed. Regarding recall, the results vary across datasets. Both experimental settings achieve the same recall on D1, \textit{DRILS-w/o-R} performs better on D2, and \textit{DRILS} performs better on D3. In fact, the average recall across the three datasets is relatively similar for the two settings, reaching 85.2\% for \textit{DRILS-w/o-R} and 87.1\% for \textit{DRILS}. The main difference lies in precision, which averages 63.53\% for \textit{DRILS-w/o-R} and 76.3\% for \textit{DRILS}. The difference is especially notable for D2, where precision drops from 86.2\% to 58.6\% without the rationale. These results suggest that the rationale plays a particularly critical role in guiding the \textit{Judge Agent} toward more precise impact set selection. Without access to the justifications provided by the \textit{Generator Agents}, the \textit{Judge Agent} lacks insight into why certain requirements were identified as impacted and may infer a different rationale, potentially leading it to incorrectly assess the generator's decision as illogical. In contrast, providing the rationale gives the \textit{Judge Agent} additional evidence to correctly and critically assess whether the proposed selection is logically justified.

Regarding operational metrics, effectiveness is similar across the two settings, though slightly higher when the rationale is included. It reached 89.7\% for \textit{DRILS} compared with 86.8\% for \textit{DRILS-w/o-R}. In contrast, the cost is consistently lower when the rationale is included. This lower cost, combined with the higher precision, suggests that providing the rationale enables the \textit{Judge Agent} to make more accurate selections while requiring fewer resources.

Overall, although the magnitude of the differences varies across metrics and datasets, \textit{DRILS} achieves higher average $F_1$ score, recall, precision, and effectiveness, while requiring a lower cost than \textit{DRILS-w/o-R}. These results indicate that the rationale is an important element of the \textit{Generator Agents}' output and provides valuable information to the \textit{Judge Agent}, particularly by improving the precision of impact set selection.

\subsubsection{\textbf{RQ1.2. Effect of Change Location on the Judge Agent}}
The results of removing the change location are captured in \autoref{tab:judge-information} under the \textit{DRILS-w/o-CL} setting. Removing the change location results in an average $F_1$ score of 83.1\% across datasets, which is slightly higher than the 80.9\% achieved by \textit{DRILS}. However, a closer look at the individual metrics shows that this $F_1$ score improvement is driven primarily by higher precision, while average recall declines from 87.1\% to 84.1\%. This suggests that without the change location, the \textit{Judge Agent} becomes more conservative in its selections, identifying a smaller but more precise set of impacted requirements. This may be because the impact of a change can sometimes be implicit in the requirement text and it becomes apparent only when the change location is explicitly provided to the \textit{Judge Agent}. While this reduces false positives, it also causes the system to miss actual impacted requirements.  Therefore, although removing change location appears beneficial for the $F_1$ score, the decline in recall suggests this information plays an important role in predicting a complete identified impact set. Given that missing impacted requirements can be more detrimental to CIA than including a small number of additional requirements, recall is prioritized over precision for this task. Thus, the change location is considered an important element of auxiliary information.

In terms of effectiveness, \textit{DRILS} achieved a higher average across the three datasets, reaching 89.7\%, compared with 87.6\% for \textit{DRILS-w/o-CL}. In contrast, \textit{DRILS-w/o-CL} requires lower cost compared to \textit{DRILS} on two of the three datasets. Specifically, \textit{DRILS-w/o-CL} achieves 4.1\%, 7.3\% and 5.3\%, compared with 5.1\%, 6.9\%, and 6.2\% for \textit{DRILS}, for D1, D2, and D3, respectively. The overall cost increase is due to the lower precision caused by including the change location. 

Overall, given the higher recall associated with including the change location, we consider it a relevant element of the \textit{Judge Agent}'s input, particularly for improving the completeness of the identified impact set.

\subsubsection{\textbf{RQ1.3. Effect of Confidence Score on the Judge Agent}}
The results of \textit{DRILS-w/o-CS} across all datasets are presented in \autoref{tab:judge-information}. The average $F_1$ scores for \textit{DRILS-w/o-CS} and \textit{DRILS} are 79.8\% and 80.9\%, respectively, reflecting a slight decline in overall performance upon removal of the confidence score. A closer examination of the results of individual datasets reveals that the effect of removing the confidence score is different across datasets. Specifically, when the confidence score is removed, recall remains the same on D1, increases by 1.7 percentage points on D2, and decreases by 7.4 percentage points on D3. Furthermore, precision decreases by 13.3 percentage points on D2, and increases by 4.2 and 7.6 percentage points on D1 and D3, respectively. These results suggest that the effect of including the confidence score is more pronounced when its effect is positive, i.e., when \textit{DRILS} achieves higher performance than \textit{DRILS-w/o-CS}. In contrast, when \textit{DRILS-w/o-CS} performs better, the difference between the two settings is smaller, suggesting that the negative effect of including the confidence score is limited. Removing the confidence score slightly reduces both average recall and average precision, with recall decreasing from 76.3\% to 75.8\% and precision from 87.1\% to 85.2\%. Among the auxiliary information elements considered, the confidence score has the smallest overall effect on the performance metrics.

Unlike the rationale and change location, the confidence score is a single integer rather than a chunk of text. Nevertheless, beyond its relatively small influence on the \textit{Judge Agent}'s decisions, it serves as an implicit quality signal for the other pieces of auxiliary information produced by the \textit{Generator Agents}. Specifically, a low confidence score may indicate that the accompanying rationale or change location is less reliable, providing the \textit{Judge Agent} with a means of weighing the trustworthiness of the information it receives. The confidence score therefore enriches the overall information context available to the \textit{Judge Agent}, enabling more calibrated decision-making. Removing it leaves the \textit{Judge Agent} without information about the reliability of other auxiliary information, which may explain the slight performance degradation observed across datasets.

Regarding operational metrics, effectiveness varies across datasets, with both settings achieving the same effectiveness on D1, \textit{DRILS-w/o-CS} achieving a higher score on D2, and \textit{DRILS} achieving a higher score on D3. When averaging across datasets, \textit{DRILS} achieves a higher score of 89.7\% than 86.8\% for \textit{DRILS-w/o-CS}, with effectiveness exceeding 90\% on two datasets. This difference is more pronounced than for the other metrics and may be because the confidence score indicates how confident each \textit{Generator Agent} is in its annotation. This information can help the \textit{Judge Agent} make a more informed decision when choosing between annotations in cases where the \textit{Generator Agents} disagree, thereby supporting a more effective decision-making process. Regarding cost, both settings have an average cost of 6\% across datasets. However, \textit{DRILS} is more consistent across datasets, ranging from 5.1\% to 6.9\%, compared with 4.7\% to 8.4\% for \textit{DRILS-w/o-CS}.

Overall, while the results vary across metrics and datasets and the differences in performance are generally small, the slight improvement in the average metrics when the confidence score is included, together with its usefulness in facilitating the analysis of the outputs, makes it a relevant element of auxiliary information to provide as input to the \textit{Judge Agent}.

\subsubsection{\textbf{RQ1.4. Effect of Combined Auxiliary Information on the Judge Agent}}
In this final setting, all three types of auxiliary information (i.e., the rationale, change location, and confidence score) are simultaneously removed from the \textit{Judge Agent}'s input. The \textit{DRILS-w/o-(R-CL-CS)} setting represents a scenario in which the \textit{Judge Agent} is provided only with the annotations produced by the \textit{Generator Agents}, without information about why or how a requirement is affected by the change request or how confident the \textit{Generator Agents} are in their annotations. As shown in \autoref{tab:judge-information}, the \textit{DRILS-w/o-(R-CL-CS)} setting achieves the lowest $F_1$ scores and recall values compared with all other settings across all three datasets, with an average $F_1$ score and recall of 70.2\% and 71.6\%, compared with 80.9\% and 87.1\% for \textit{DRILS}. This corresponds to relative decreases of 13.12\% and 17.80\% in $F_1$ score and recall, respectively, representing the largest performance declines observed across all settings. Average precision is also lower for \textit{DRILS-w/o-(R-CL-CS)} at 69.7\% compared with 76.3\% for \textit{DRILS}. Effectiveness also decreases, with \textit{DRILS-w/o-(R-CL-CS)} achieving the lowest average of 79.8\%, compared with 89.7\% for \textit{DRILS}. In contrast, the two settings have similar costs, averaging about 6\% each.

These results indicate that while each individual element of auxiliary information may have a moderate effect on the \textit{Judge Agent}'s decision-making, their combined presence has a substantially greater effect on almost all evaluated metrics. The substantial performance drop when removing all auxiliary information suggests that the rationale, change location, and confidence score complement one another, and that their combined presence is essential for the \textit{Judge Agent} to accurately identify the final impact set.

\subsubsection{\textbf{Robustness Analysis}}

To ensure the robustness of the results, we ran each configuration three times and report the deviation across runs. As \autoref{tab:judge-information} shows, the average of the deviations across all settings and datasets is 0.027. Among the 75 experimental results, 21.3\% exhibit no variation across runs (i.e., a deviation of 0.00), while 88\% have a deviation below 0.05. Only a few instances exceed 0.05: 0.06 in three instances and 0.08 and 0.09 in one instance each. \textit{DRILS} demonstrates high stability across runs, with deviations below 0.05 for 14 of the 15 metric–dataset combinations and a maximum deviation of 0.05 in the remaining case. Overall, these results indicate low variability across repeated executions for most configurations, suggesting that the obtained results are robust despite the inherent stochasticity of LLM-based approaches.

\begin{tcolorbox}[arc=1mm,width=\columnwidth,
                  top=0mm,left=0mm,  right=0mm, bottom=0mm,
                  boxrule=1pt, colback=blue!15!white,colframe=white]
\textbf{Answer to RQ1:} While the effects of the auxiliary information generated by the \textit{Generator Agents}, namely the Rationale, Change Location, and Confidence Score, are moderate when evaluated individually, their combined contribution is much more pronounced in guiding the \textit{Judge Agent} during impact set identification. When all auxiliary information is included, \textit{DRILS} achieves an average $F_1$ score of 80.9\%, a recall of 87.1\%, a precision of 76.3\%, and an effectiveness of 89.7\%. The Rationale appears to guide the \textit{Judge Agent} toward more precise selections, while the Change Location contributes to a more effective identification of impacted requirements. \end{tcolorbox} \subsection{RQ2: Effect of pre-processing stages on \textit{DRILS} performance}\label{sec:RQ2}
\begin{table}
\centering
\centering
\caption{Comparison of DRILS, DRILS-w/o-RE-RAG, and DRILS-RAG}
\label{tab:rq2-srag}
\begin{tabularx}{\linewidth}{
    @{\hskip 0.1em}
    >{\raggedright\arraybackslash}m{0.05\textwidth}
    >{\raggedright\arraybackslash}m{0.05\textwidth}
    @{\hskip 0.5em}
    *{3}{>{\centering\arraybackslash}X}
    @{\hskip 0.1em}
}
    \toprule
    \textbf{Dataset} & \textbf{Metric} & \textbf{DRILS-w/o-RE-RAG} & \textbf{DRILS-RAG} & \textbf{DRILS} \\
    \midrule
    \multirow{5}{*}{\textbf{D1}}
    & \textbf{R}    & $81.9 \pm 0.08$ & $88.8 \pm 0.05$ & $\mathbf{91.6 \pm 0.00}$ \\
    & \textbf{P}    & $62.8 \pm 0.14$ & $\mathbf{66.1 \pm 0.03}$ & $64.9 \pm 0.05$ \\
    & \textbf{F1}   & $70.8 \pm 0.10$ & $75.7 \pm 0.01$ & $\mathbf{75.9 \pm 0.03}$ \\
    & \textbf{Eff}  & $87.2 \pm 0.06$ & $91.6 \pm 0.05$ & $\mathbf{94.9 \pm 0.00}$ \\
    & \textbf{Cost} & $\mathbf{04.8 \pm 0.14}$ & $04.9 \pm 0.08$ & $05.1 \pm 0.05$ \\
    \midrule
    \multirow{5}{*}{\textbf{D2}}
    & \textbf{R}    & $83.3 \pm 0.06$ & $81.6 \pm 0.03$ & $\mathbf{83.3 \pm 0.03}$ \\
    & \textbf{P}    & $\mathbf{89.2 \pm 0.01}$ & $87.8 \pm 0.08$ & $86.2 \pm 0.03$ \\
    & \textbf{F1}   & $\mathbf{86.1 \pm 0.04}$ & $84.5 \pm 0.05$ & $84.8 \pm 0.03$ \\
    & \textbf{Eff}  & $82.3 \pm 0.04$ & $81.4 \pm 0.02$ & $\mathbf{82.3 \pm 0.01}$ \\
    & \textbf{Cost} & $\mathbf{06.6 \pm 0.06}$ & $06.6 \pm 0.08$ & $06.9 \pm 0.03$ \\
    \midrule
    \multirow{5}{*}{\textbf{D3}}
    & \textbf{R}    & $75.3 \pm 0.03$ & $75.3 \pm 0.06$ & $\mathbf{86.4 \pm 0.02}$ \\
    & \textbf{P}    & $77.7 \pm 0.11$ & $73.5 \pm 0.11$ & $\mathbf{77.8 \pm 0.04}$ \\
    & \textbf{F1}   & $76.4 \pm 0.07$ & $74.1 \pm 0.03$ & $\mathbf{81.9 \pm 0.03}$ \\
    & \textbf{Eff}  & $82.9 \pm 0.07$ & $84.4 \pm 0.03$ & $\mathbf{91.7 \pm 0.01}$ \\
    & \textbf{Cost} & $\mathbf{05.4 \pm 0.08}$ & $05.8 \pm 0.16$ & $06.2 \pm 0.03$ \\
    \bottomrule
\end{tabularx}
\end{table}
 
To evaluate the effects of \textit{DRILS}’s two pre-processing steps, RE-RAG and Clustering (steps 1 and 2 in the \textit{DRILS} architecture presented in \autoref{fig:methodology} and explained in Section~\ref{sec:approach}), we assess different settings by systematically replacing each step with alternative techniques. We compute recall, precision, $F_1$ score, effectiveness, and cost to evaluate overall performance across all settings, as well as the performance of each Generator Agent individually and their union (i.e., using the combined set of requirements identified as impacted by both agents).

\subsubsection{\textbf{RQ2.1. Comparison of RE-RAG with No RAG and Standard RAG}}
\label{sec:rq2-srag}

To evaluate the effect of RE-RAG on the overall performance of \textit{DRILS}, we assess two alternative techniques: (1) \textit{DRILS-w/o-RE-RAG}, where RAG is not used within \textit{DRILS}, i.e., only the original requirements are provided to the agents without additional domain-specific context, and (2) \textit{DRILS-RAG}, where a standard chunk-based RAG is used. In the latter setting, the SRS documents are divided into text chunks of 1,000 tokens with an overlap of 200 tokens. These chunks are then analyzed based on their semantic similarity to the change request using the Cosine similarity metric, and the three most similar chunks are 
provided to the LLM as contextual information. The prompt used by the \textit{Generator Agents} (\autoref{fig:main-prompt}) remains unchanged, except for the removal of RE-RAG-related instructions and the inclusion of the retrieved chunks in the input section of the prompt (\textbf{\#\# Input} in \autoref{fig:main-prompt}).

\paragraph{\textbf{Overall Performance}}
Table~\ref{tab:rq2-srag} presents the results of the two alternative settings alongside \textit{DRILS} with RE-RAG. The average $F_1$ score across all datasets for \textit{DRILS-w/o-RE-RAG}, \textit{DRILS-RAG}, and \textit{DRILS} are 77.6\%, 78.1\%, and 80.9\%, respectively. This indicates that incorporating RE-RAG slightly improves overall performance, with relative $F_1$ score improvements of 4.3\% and 3.6\% for \textit{DRILS-w/o-RE-RAG} and \textit{DRILS-RAG}, respectively. While the difference is relatively small for $F_1$ score, it is more pronounced for recall. \textit{DRILS} with RE-RAG achieves the highest recall across all datasets, with scores above 80\% on each. The largest recall gain is observed for D3, where recall increases by more than 10 percentage points. However, in D2, \textit{DRILS-w/o-RE-RAG} matches the recall of \textit{DRILS} with RE-RAG but is less stable, with a deviation of 0.06 compared with 0.03 for \textit{DRILS}. Regarding precision, no single setting consistently achieves the highest score, as each setting obtains the highest precision on one of the three datasets. Moreover, the average precision is nearly identical across the three settings, at approximately 76\%. Overall, since we prioritize recall over $F_1$ score and precision for the CIA task, these results indicate that incorporating RE-RAG is beneficial for \textit{DRILS}.

These results indicate that RE-RAG provides domain-specific information that helps the agents identify more relevant requirements rather than improving precision. This might be because domain knowledge enables the identification of implicit links between the change request and the requirements that might not be evident from the requirement text alone. Furthermore, providing more specific information and pointers for resolving referential ambiguities, along with domain knowledge in concise bullet points, may reduce noise and irrelevant information that could confuse the LLM. In fact, \textit{DRILS-w/o-RE-RAG} and \textit{DRILS-RAG} achieve very similar results across the average metrics, with differences of less than one percentage point for $F_1$ score and precision, and less than two percentage points for recall (77.6\% and 78.1\% for $F_1$ score, 80.1\% and 81.9\% for recall, and 76.5\% and 75.8\% for precision). This suggests that standard RAG without the enhancements provided by RE-RAG has limited impact on the CIA task.

Effectiveness is highest and most stable when RE-RAG is used, outperforming both alternative settings. It exceeds 90\% on two of the three datasets, a level reached only once in the other two settings. On average, \textit{DRILS} achieves an effectiveness of 89.6\%, compared with 84.1\% for \textit{DRILS-w/o-RE-RAG} and 85.8\% for \textit{DRILS-RAG}. In terms of cost, \textit{DRILS-w/o-RE-RAG} is the cheapest, although the difference is relatively small. These results indicate that while incorporating RAG (chunk-based or RE-RAG) improves effectiveness, it also requires more cost. The increase in effectiveness may result from the additional context provided by RAG, which helps identify a broader set of potentially impacted requirements and thus contributes to a more complete impact set. However, retrieving a broader set of requirements also increases the proportion of requirements that must be manually validated, resulting in higher cost.

\paragraph{\textbf{Robustness Analysis}} As \autoref{tab:rq2-srag} shows, the robustness results vary across settings. The average deviation is 0.07, 0.05, and 0.02 for \textit{DRILS-w/o-RE-RAG}, \textit{DRILS-RAG}, and \textit{DRILS}, respectively. In fact, \textit{DRILS-w/o-RE-RAG} shows relatively high variability, exceeding 0.05 on 11 out of 15 dataset-metric combinations, with a maximum deviation of 0.14 for precision on D1. \textit{DRILS-RAG} exceeds this threshold in 6 instances, with a maximum deviation of 0.16 for cost on D3. In contrast, the deviation of \textit{DRILS} is equal to or below 0.05 for all dataset-metric combinations. These results suggest that incorporating additional context through RAG can reduce the variability across runs. This may be because the additional context gives agents information that effectively guides their responses. This is more pronounced when the context is precisely tailored to the change request, as in RE-RAG. By providing only relevant contextual information, RE-RAG may reduce noise, thereby leading to more consistent results across runs.

\begin{table*}
\centering
\caption{Generator Agents Stage Analysis - RE-RAG}
\label{tab:generator-performance}
\begin{tabularx}{\textwidth}{
    @{\hskip 0.5em}
    >{\centering\arraybackslash}m{0.1\textwidth}
    @{\hskip 0.5em}
    >{\raggedright\arraybackslash}m{0.20\textwidth}
    @{\hskip 0.5em}
    *{9}{>{\centering\arraybackslash}X}
    @{\hskip 0.5em}
}
    \toprule
    \multirow{3}{*}{\textbf{Category}} &
    \multirow{3}{*}{\textbf{Method}} &
    \multicolumn{3}{c}{\textbf{D1}} &
    \multicolumn{3}{c}{\textbf{D2}} &
    \multicolumn{3}{c}{\textbf{D3}} \\
    
    \cmidrule(lr){3-5}
    \cmidrule(lr){6-8}
    \cmidrule(lr){9-11}
    
    & &
    \textbf{R} & \textbf{P} & \textbf{F1} &
    \textbf{R} & \textbf{P} & \textbf{F1} &
    \textbf{R} & \textbf{P} & \textbf{F1} \\
    
    \midrule
    \multirow{3}{*}{Agent1}  & DRILS-w/o-RE-RAG    & 81.9 & 23.1 & 36.1 & \textbf{93.3} & \textbf{27.6} & \textbf{42.6} & \textbf{91.4} & 19.8 & 32.6 \\
                         & DRILS-RAG         & \textbf{87.5} & \textbf{23.4} & \textbf{36.8} & 83.3 & 23.6 & 36.9 & 90.1 & 19.0 & 31.3 \\
                         & DRILS             & 84.7 & 18.0 & 29.7 & 60.0 & 20.3 & 30.3 & 84.0 & \textbf{21.4} & \textbf{34.1} \\
\midrule
\multirow{3}{*}{Agent2} & DRILS-w/o-RE-RAG    & 88.9 & 35.4 & 51.1 & \textbf{96.7} & 27.6 & 43.0 & \textbf{93.8} & 32.6 & 48.6 \\
                        & DRILS-RAG         & 87.5 & 32.4 & 47.4 & 85.0 & 27.8 & 42.4 & 88.9 & 28.4 & 43.1 \\
                        & DRILS             & \textbf{93.1} & \textbf{39.4} & \textbf{57.0} & 81.7 & \textbf{36.4} & \textbf{51.9} & 85.2 & \textbf{46.0} & \textbf{59.8} \\
\midrule
\multirow{3}{*}{Union} & DRILS-w/o-RE-RAG    & 91.7 & \textbf{21.7} & \textbf{35.2} & \textbf{96.7} & \textbf{22.9} & \textbf{36.9} & \textbf{95.1} & 17.6 & 29.5 \\
                       & DRILS-RAG         & 93.1 & 21.3 & 34.6 & \textbf{96.7} & 22.2 & 36.0 & 93.8 & 17.0 & 28.8 \\
                       & DRILS             & \textbf{95.8} & 18.8 & 31.5 & 88.3 & 22.6 & 36.2 & 89.0 & \textbf{19.9} & \textbf{32.6} \\
    \bottomrule
\end{tabularx}

\end{table*}

\paragraph{\textbf{Analysis of the Behavior of Generator Agents}}

To assess whether the performance difference stems from the \textit{Generator Agents} or the \textit{Judge Agent}, we analyze the \textit{Generator Agents}' performance and compare it to the overall performance. \autoref{tab:generator-performance} presents the precision, recall, and $F_1$ score for each \textit{Generator Agent}, as well as for their union, for all settings and datasets. As the table shows, the average $F_1$ score for \textit{DRILS-w/o-RE-RAG} is 37.1\% for Agent1, 47.5\% for Agent2, and 33.8\% for the union, while for \textit{DRILS-RAG}, it is 35.0\% for Agent1, 44.3\% for Agent2, and 33.1\% for the union. Finally, \textit{DRILS}, averages 31.3\% for Agent1, 56.2\% for Agent2, and 33.4\% for the union. \textit{DRILS-w/o-RE-RAG} has the highest $F_1$ score for Agent1 and the union, while \textit{DRILS} has the highest average $F_1$ score for Agent2. 
The low $F_1$ scores are due to low precision, which ranged from 18.0\% to 46.0\% for individual agents and from 17.0\% to 22.9\% for the union. These precision scores are extremely low, given that we did not instruct the agents to be as inclusive as possible. The high number of false positives can introduce noise and potentially confuse the \textit{Judge Agent}, while increasing its filtering workload.
Nevertheless, because the \textit{Generator Agents} are followed by a \textit{Judge Agent} that filters, this low precision does not harm the overall architecture.  Overall, the \textit{Generator Agents} are not precise in their selection, leaving the \textit{Judge Agent} to filter false-positive candidates.

Regarding recall, the \textit{Generator Agents} generally achieve high recall across settings and datasets. For \textit{DRILS-w/o-RE-RAG}, it averaged 88.8\% for Agent1, 93.1\% for Agent2, and 94.5\% for the union, while for \textit{DRILS-RAG}, it averaged 86.9\% for Agent1, 87.1\% for Agent2, and 94.5\% for the union, and for  for \textit{DRILS}, it averaged 76.2\% for Agent1, 86.6\% for Agent2, and 91.0\% for the union. \textit{DRILS-w/o-RE-RAG} achieves the highest average recall for Agent1, Agent2, and the union. In fact, \textit{DRILS} has the lowest average recall across all three settings, possibly because it retrieves fewer requirements than the other two settings. These results suggest that the additional contextual information provided by RE-RAG makes the Generator Agents more selective in identifying potentially impacted requirements. Without RE-RAG, the agents rely primarily on the requirements' text alone and may identify a broader set of potentially related requirements, resulting in higher recall.

In our context, recall of the union is particularly important, as the \textit{Judge Agent} can select only impacted requirements from the candidates provided by the \textit{Generator Agents}. Thus, it is necessary to ensure that the \textit{Judge Agent} is provided with as many potentially impacted requirements as possible. In this regard, \textit{DRILS-w/o-RE-RAG} achieves the highest union recall and is therefore considered the best-performing setting for the \textit{Generator Agents}. However, the recall differences between each individual agent and their union are 21 percentage points for \textit{DRILS-w/o-RE-RAG}, 44.9 percentage points for \textit{DRILS-RAG}, and 57.5 percentage points for \textit{DRILS} across datasets. A larger difference suggests that the agents identify more diverse sets of requirements. Thus, for \textit{DRILS}, the agents identify more diverse sets of requirements, resulting in broad collective coverage despite lower individual recall than in the other settings. This also suggests that the agents in the other settings tend to identify a larger number of the same true-positive requirements.

When comparing the union recall of the agents with that of the \textit{Judge Agent}, a different pattern emerges. Although \textit{DRILS} achieved the lowest average union recall for the \textit{Generator Agents}, it achieved the highest overall average recall. The average recall dropped only from 91.0\% to 87.1\% (a 3.9 percentage point drop), compared with \textit{DRILS-w/o-RE-RAG}, which dropped from 94.5\% to 80.0\% (a 14.5 percentage point drop). Thus, although removing RAG improved the performance of the \textit{Generator Agents}, both individually and in union, it reduced final recall. This may be due to the lower quality of the auxiliary information than in \textit{DRILS}. Because the \textit{Generator Agents} lack domain context, their rationale may be less accurate and therefore less useful to the \textit{Judge Agent}. In contrast, for \textit{DRILS}, although the \textit{Generator Agents} identified fewer true positives, the \textit{Judge Agent} correctly identified more impacted requirements from that pool, likely due to the more grounded auxiliary information. Also, for \textit{DRILS-w/o-RE-RAG}, the overall recall is lower than that of Agent2 (the best-performing agent) for all three datasets. This suggests that the \textit{Judge Agent} does not rely more on the predictions of the best-performing \textit{Generator Agent}. This might be because Agent2 has a lower average confidence score for true positives than Agent1 (e.g., 4.62 vs. 4.95 for D1, and 4.34 vs. 4.71 for D3), suggesting Agent2 is less confident in its decisions and thus less influential in the Judge's decision. In contrast, for \textit{DRILS}, the overall recall is higher than that of Agent2 for two out of the three datasets. While it does not indicate that the Judge took all the responses given by Agent2 into consideration, it suggests that the Judge was able to make more effective use of the \textit{Generator Agents}' outputs compared to \textit{DRILS-w/o-RE-RAG}.

A common pattern across all these settings is that Agent2 performs significantly better than Agent1 for the task. This is reflected in consistently higher precision and recall across almost all configurations, indicating that Agent2 (Claude) is more effective for the task.
This difference is particularly pronounced for \textit{DRILS}, which achieved the lowest average $F_1$ score for Agent1 and the highest average $F_1$ score for Agent2. This difference may be attributed to the different capabilities of the agents in analyzing and leveraging the enriched contextual information provided by RE-RAG.

Overall, while excluding RE-RAG generally resulted in higher recall for the \textit{Generator Agents}, including it provided a higher overall output recall. These results indicate that RE-RAG has mainly affected the \textit{Judge Agent}'s decision-making rather than the \textit{Generator Agents}' impact set identification. This may be due to the superior quality of the auxiliary information, as it provides more relevant and focused domain context.

\begin{table}
\centering
\centering
\caption{Comparison of DRILS, DRILS-w/o-Cluster, and DRILS-Batch.}
\label{tab:rq2-clustering}
\begin{tabularx}{\linewidth}{
    @{\hskip 0.1em}
    >{\raggedright\arraybackslash}m{0.05\textwidth}
    >{\raggedright\arraybackslash}m{0.05\textwidth}
    @{\hskip 0.5em}
    *{3}{>{\centering\arraybackslash}X}
    @{\hskip 0.1em}
}
    \toprule
    \textbf{Dataset} & \textbf{Metric} & \textbf{DRILS-w/o-Cluster} & \textbf{DRILS-Batch} & \textbf{DRILS} \\
    \midrule
    \multirow{5}{*}{\textbf{D1}}
    & \textbf{R}    & $81.9 \pm 0.02$ & $87.4 \pm 0.05$ & $\mathbf{91.6 \pm 0.00}$ \\
    & \textbf{P}    & $73.1 \pm 0.08$ & $\mathbf{76.2 \pm 0.06}$ & $64.9 \pm 0.05$ \\
    & \textbf{F1}   & $77.2 \pm 0.05$ & $\mathbf{81.3 \pm 0.03}$ & $75.9 \pm 0.03$ \\
    & \textbf{Eff}  & $84.9 \pm 0.03$ & $92.1 \pm 0.04$ & $\mathbf{94.9 \pm 0.00}$ \\
    & \textbf{Cost} & $\mathbf{04.0 \pm 0.07}$ & $04.2 \pm 0.11$ & $05.1 \pm 0.05$ \\
    \midrule
    \multirow{5}{*}{\textbf{D2}}
    & \textbf{R}    & $\mathbf{85.0 \pm 0.05}$ & $83.3 \pm 0.03$ & $83.3 \pm 0.03$ \\
    & \textbf{P}    & $\mathbf{92.6 \pm 0.02}$ & $83.8 \pm 0.08$ & $86.2 \pm 0.03$ \\
    & \textbf{F1}   & $\mathbf{88.6 \pm 0.02}$ & $83.1 \pm 0.05$ & $84.8 \pm 0.03$ \\
    & \textbf{Eff}  & $\mathbf{83.3 \pm 0.03}$ & $82.5 \pm 0.02$ & $82.5 \pm 0.01$ \\
    & \textbf{Cost} & $\mathbf{06.6 \pm 0.08}$ & $07.3 \pm 0.08$ & $06.9 \pm 0.03$ \\
    \midrule
    \multirow{5}{*}{\textbf{D3}}
    & \textbf{R}    & $80.0 \pm 0.02$ & $81.3 \pm 0.03$ & $\mathbf{86.4 \pm 0.02}$ \\
    & \textbf{P}    & $\mathbf{84.2 \pm 0.01}$ & $79.3 \pm 0.01$ & $77.8 \pm 0.04$ \\
    & \textbf{F1}   & $81.5 \pm 0.02$ & $79.7 \pm 0.01$ & $\mathbf{81.9 \pm 0.03}$ \\
    & \textbf{Eff}  & $87.1 \pm 0.06$ & $89.1 \pm 0.02$ & $\mathbf{91.7 \pm 0.01}$ \\
    & \textbf{Cost} & $\mathbf{05.2 \pm 0.02}$ & $05.7 \pm 0.04$ & $06.2 \pm 0.03$ \\
    \bottomrule
\end{tabularx}
\end{table}
 \subsubsection{\textbf{RQ2.2. Comparison of Clustering with Full List and Fixed Batching}}
\label{sec:rq2-clustering}

In this section, we analyze the results of two alternative techniques to the clustering step within \textit{DRILS}: (1) \textit{DRILS-w/o-Cluster} where all requirements are provided at once and no technique is applied to divide the requirements list before providing it to the \textit{Generator Agents}, and (2) \textit{DRILS-Batch}, where the requirements list is divided into fixed-size batches. For the latter, we split each dataset's requirements list into two batches. This batch size was chosen to minimize the cost of API calls for both the GPT and Claude models. In \textit{DRILS}, the \textit{Generator Agents} processed each cluster individually, which corresponds to three, two, and three clusters for D1, D2, and D3, respectively.

\paragraph{\textbf{Overall Performance.}}
Table~\ref{tab:rq2-clustering} presents results for these two settings alongside \textit{DRILS}. For $F_1$ score, no setting is clearly superior, as each achieves the highest score on one dataset: \textit{DRILS-Batch} on D1, \textit{DRILS-w/o-Cluster} on D2, and \textit{DRILS} on D3. 
The higher $F_1$ scores of \textit{DRILS-w/o-cluster} and \textit{DRILS-Batch} compared with \textit{DRILS} are driven by their higher precision. The average precision reached 83.3\%, 79.7\%, and 76.3\% for \textit{DRILS-w/o-Cluster}, \textit{DRILS-Batch}, and \textit{DRILS}, respectively. The higher precision when the requirements list is not divided may be due to the broader context it provides to the \textit{Generator Agents}, which may help them better distinguish between impacted and non-impacted requirements. This effect is particularly pronounced for D2, where \textit{DRILS-w/o-Cluster} achieves the highest performance across all metrics. However, D2 has fewer requirements than APA and ADB, which likely explains this behavior. When the total number of requirements is small, providing them all in a single prompt gives the model complete context without encountering the lost-in-the-middle problem~\cite{liu2024lost}.

In contrast, dividing the requirements list provides an advantage in terms of recall. \textit{DRILS} achieves the highest average recall of 87.1\%, compared with 82.3\% for \textit{DRILS-w/o-Cluster} and 84\%. This indicates that dividing the requirements list allows the identification of more potentially impacted requirements. This may be because splitting a large requirements list into smaller sets reduces the information the \textit{Generator Agents} must process simultaneously, allowing them to focus more effectively on the requirements relevant to the change request. This is particularly important for safety-critical systems, where the number of requirements can be substantially larger than the evaluated datasets. In this case, injecting all requirements into a single prompt becomes impractical and non-scalable. When comparing \textit{DRILS-Batch} with \textit{DRILS}, \textit{DRILS} achieves higher recall at the cost of a marginal decrease in precision, making it the more suitable choice for the CIA task. This may be because clustering divides requirements more effectively by grouping similar and related requirements into the same cluster, giving the \textit{Generator Agents} more relevant context than simple size-based batching.

Regarding effectiveness, \textit{DRILS} achieves higher average effectiveness than the other two settings, surpassing 90\% on two of the three datasets. In terms of cost, \textit{DRILS-w/o-Cluster} has the lowest cost, followed by \textit{DRILS-Batch} and \textit{DRILS}. These results indicate that while clustering achieves the highest effectiveness, it also incurs the highest cost, likely because it retrieves larger impact sets for validation, requiring the requirements analyst to review a greater proportion of requirements.

\paragraph{\textbf{Robustness Analysis}} Overall, the results are robust, with an average deviation of 0.033 across the three settings. Specifically, the average deviation is 0.038 for \textit{DRILS-w/o-cluster}, 0.037 for \textit{DRILS-Batch}, and 0.026 for \textit{DRILS}. These results indicate that dividing the requirements list, particularly through a similarity-based strategy rather than fixed-size batching, reduces result variability across runs. This may be because the \textit{Generator Agents} need to process less information simultaneously, allowing them to focus on a smaller, more coherent subset of requirements.

\begin{table*}
\centering
\caption{Generator Agents Stage Analysis - Clustering}
\label{tab:generator-cluster}
\begin{tabularx}{\textwidth}{
    @{\hskip 0.5em}
    >{\centering\arraybackslash}m{0.1\textwidth}
    @{\hskip 0.5em}
    >{\raggedright\arraybackslash}m{0.20\textwidth}
    @{\hskip 0.5em}
    *{9}{>{\centering\arraybackslash}X}
    @{\hskip 0.5em}
}
    \toprule
    \multirow{3}{*}{\textbf{Category}} &
    \multirow{3}{*}{\textbf{Method}} &
    \multicolumn{3}{c}{\textbf{D1}} &
    \multicolumn{3}{c}{\textbf{D2}} &
    \multicolumn{3}{c}{\textbf{D3}} \\
    
    \cmidrule(lr){3-5}
    \cmidrule(lr){6-8}
    \cmidrule(lr){9-11}
    
    & &
    \textbf{R} & \textbf{P} & \textbf{F1} &
    \textbf{R} & \textbf{P} & \textbf{F1} &
    \textbf{R} & \textbf{P} & \textbf{F1} \\
    
    \midrule
    \multirow{3}{*}{Agent1}  & DRILS-w/o-Cluster & 59.7 & \textbf{37.7} & \textbf{46.2} & \textbf{90.0} & \textbf{37.1} & \textbf{52.5} & \textbf{90.1} & \textbf{35.4} & \textbf{51.1} \\
                         & DRILS-Batch       & 68.1 & 24.3 & 35.7 & 60.0 & 26.5 & 37.8 & 81.5 & 34.1 & 48.1 \\
                         & DRILS             & \textbf{84.7} & 18.0 & 29.7 & 60.0 & 20.3 & 30.3 & 84.0 & 21.4 & 34.1 \\
\midrule
\multirow{3}{*}{Agent2} & DRILS-w/o-Cluster  & 88.9 & 38.4 & 53.9 & 86.7 & \textbf{56.5} & \textbf{68.5} & 77.8 & \textbf{71.7} & \textbf{74.6} \\
                        & DRILS-Batch        & 88.9 & \textbf{43.3} & \textbf{58.3} & \textbf{91.7} & 39.3 & 55.6 & \textbf{85.2} & 51.4 & 63.9 \\
                        & DRILS              & \textbf{93.1} & 39.4 & 57.0 & 81.7 & 36.4 & 51.9 & \textbf{85.2} & 46.0 & 59.8 \\
\midrule
\multirow{3}{*}{Union} & DRILS-w/o-Cluster   & 88.9 & \textbf{38.4} & \textbf{53.9} & 90.0 & \textbf{37.1} & \textbf{52.5} & 90.1 & \textbf{35.4} & 51.1 \\
                       & DRILS-Batch         & 88.9 & 24.8 & 38.7 & \textbf{95.0} & 30.0 & 45.6 & \textbf{93.8} & 34.3 & \textbf{51.5} \\
                       & DRILS               & \textbf{95.8} & 18.8 & 31.5 & 88.3 & 22.6 & 36.2 & 89.0 & 19.9 & 32.6 \\
    \bottomrule
\end{tabularx}

\end{table*}

\paragraph{\textbf{Analysis of the Behavior of Generator Agents}}

\autoref{tab:generator-cluster} presents the performance metrics for each \textit{Generator Agent} and their union across the three settings and datasets. The average $F_1$ scores for Agent1, Agent2, and their union for \textit{DRILS-w/o-Cluster} are 49.9\%, 65.7\%, and 52.5\%, respectively. For \textit{DRILS-Batch}, the corresponding average $F_1$ scores are 40.5\%, 59.3\%, and 45.3\%, while for \textit{DRILS}, they reach 31.3\%, 56.2\%, and 33.4\%, respectively. These results indicate that \textit{DRILS-w/o-Cluster} is the best setting for the \textit{Generator Agents}, both individually and in union, possibly because they are provided with a broader context. However, because processing all requirements simultaneously can be impractical in real-world scenarios with large requirements specifications, particularly for safety-critical systems, it is important to identify which of the two grouping techniques, batching or clustering, is more effective for the \textit{Generator Agents}.

Compared with \textit{DRILS}, \textit{DRILS-Batch} achieves higher average $F_1$ scores. This is mainly due to its higher precision, which averages 34.2\% compared with 27.0\% for \textit{DRILS}. In contrast, recall is marginally higher for \textit{DRILS}, averaging 84.6\% compared with 83.6\% for \textit{DRILS-Batch}. This difference is mainly attributable to \textit{DRILS}' superior performance for Agent1 on two of the three datasets. However, \textit{DRILS-Batch} achieves marginally higher union recall (92.5\% versus 91.0\%). Since union recall is particularly important for our evaluation, as the goal of \textit{Impact-set Identification} is to provide the \textit{Judge Agent} with a complete set of impacted requirements, we consider \textit{DRILS-Batch} the most effective technique for dividing the requirements list at the \textit{Generator Agent} level.

While a high union recall indicates strong candidate coverage by the \textit{Generator Agents}, it does not guarantee that all impacted requirements will be retained in the \textit{Judge Agent}'s final output. This is evident in \autoref{tab:rq2-clustering}, where \textit{DRILS} achieves a higher average recall than both \textit{DRILS-w/o-Cluster} and \textit{DRILS-Batch}, even though the latter two settings retrieve a larger number of candidate requirements at the \textit{Generator Agents} stage. To investigate the reasons underlying this difference, we examine the average confidence scores assigned to True Positive (TP) and False Positive (FP) requirements across settings. The average confidence scores for TPs and FPs are 4.68 and 4.02 for \textit{DRILS-w/o-Cluster}, 4.69 and 3.86 for \textit{DRILS-Batch}, and 4.73 and 3.60 for \textit{DRILS}, respectively. These results indicate that \textit{DRILS} assigns the highest confidence scores to TPs and the lowest to FPs, suggesting that the \textit{Generator Agents} assign higher confidence to truly impacted requirements while being more cautious about false positives. In contrast, \textit{DRILS-w/o-Cluster} and \textit{DRILS-Batch} show a smaller difference, mainly because their FP confidence scores are higher. This smaller margin may hinder the \textit{Judge Agent}'s ability to distinguish between truly impacted and non-impacted requirements, which can explain the low recall for these two settings.

When comparing agents, Agent2 outperforms Agent1 across almost all settings and datasets, achieving higher average $F_1$ score, recall, and precision. This consistent improvement indicates that Agent2 performs better overall on the CIA task. Moreover, Agent2's performance appears to be less affected by the number of input requirements. This is reflected in the smaller difference between the $F_1$ score of \textit{DRILS-w/o-Cluster}, which processes all requirements simultaneously, and those of \textit{DRILS-Batch} and \textit{DRILS}, which process smaller groups of requirements. Specifically, when comparing the $F_1$ score of \textit{DRILS-w/o-Cluster} with the minimum $F_1$ score between \textit{DRILS-Batch} and \textit{DRILS}, Agent1 exhibits differences of 15.9, 22.2, and 17.0 percentage points across D1, D2, and D3, respectively. In contrast, Agent2's corresponding differences are much smaller, at 4.4, 16.6, and 14.8 percentage points, suggesting that Agent2 is comparatively less sensitive to changes in the number of input requirements. 

Overall, while providing the entire requirements list resulted in higher performance across all metrics for the \textit{Generator Agents}, clustering requirements yielded higher overall output recall. These results suggest that clustering primarily affects the \textit{Judge Agent}'s decision-making process rather than the \textit{Generator Agents}' identification of impact sets. This may be because the \textit{Generator Agents} assign more meaningful  confidence scores, thereby better guiding the \textit{Judge Agent}'s decisions

\begin{tcolorbox}[arc=1mm,width=\columnwidth,
                  top=0mm,left=0mm,  right=0mm, bottom=0mm,
                  boxrule=1pt, colback=blue!15!white,colframe=white]
\textbf{Answer to RQ2:} 
Integrating both RE-RAG and clustering yields higher overall recall than excluding these techniques or using alternative configurations. These pre-processing techniques primarily benefit the \textit{Judge Agent} rather than the \textit{Generator Agents}, potentially due to the higher quality of the auxiliary information they provide.
\end{tcolorbox} \subsection{RQ3: Effect of Requirement Dependency Graph on DRILS performance}\label{sec:RQ3}
\begin{table}
\centering
\caption{Comparison of DRILS and DRILS-w/o-RDG.}
\label{tab:rq3}
\begin{tabularx}{\linewidth}{
    @{\hskip 0.1em}
    >{\raggedright\arraybackslash}m{0.05\textwidth}
    >{\raggedright\arraybackslash}m{0.05\textwidth}
    @{\hskip 0.5em}
    *{2}{>{\centering\arraybackslash}X}
    @{\hskip 0.1em}
}
    \toprule
    \textbf{Dataset} & \textbf{Metric} & \textbf{DRILS-w/o-RDG} & \textbf{DRILS} \\

    \midrule
    \multirow{5}{*}{\textbf{D1}}
    & \textbf{R}    & $87.5 \pm 0.05$ & $\mathbf{91.6 \pm 0.00}$ \\
    & \textbf{P}    & $63.3 \pm 0.09$ & $\mathbf{64.9 \pm 0.05}$ \\
    & \textbf{F1}   & $73.3 \pm 0.06$ & $\mathbf{75.9 \pm 0.03}$ \\
    & \textbf{Eff}  & $89.9 \pm 0.06$ & $\mathbf{94.9 \pm 0.00}$ \\
    & \textbf{Cost} & $\mathbf{05.0 \pm 0.09}$ & $05.1 \pm 0.05$ \\

    \midrule
    \multirow{5}{*}{\textbf{D2}}
    & \textbf{R}    & $78.3 \pm 0.04$ & $\mathbf{83.3 \pm 0.03}$ \\
    & \textbf{P}    & $81.1 \pm 0.02$ & $\mathbf{86.2 \pm 0.03}$ \\
    & \textbf{F1}   & $79.6 \pm 0.01$ & $\mathbf{84.8 \pm 0.03}$ \\
    & \textbf{Eff}  & $79.6 \pm 0.02$ & $\mathbf{82.5 \pm 0.01}$ \\
    & \textbf{Cost} & $06.9 \pm 0.06$ & $\mathbf{06.9 \pm 0.03}$ \\

    \midrule
    \multirow{5}{*}{\textbf{D3}}
    & \textbf{R}    & $77.8 \pm 0.08$ & $\mathbf{86.4 \pm 0.02}$ \\
    & \textbf{P}    & $77.0 \pm 0.02$ & $\mathbf{77.8 \pm 0.04}$ \\
    & \textbf{F1}   & $77.2 \pm 0.03$ & $\mathbf{81.9 \pm 0.03}$ \\
    & \textbf{Eff}  & $87.2 \pm 0.03$ & $\mathbf{91.7 \pm 0.01}$ \\
    & \textbf{Cost} & $\mathbf{05.7 \pm 0.10}$ & $06.2 \pm 0.03$ \\
    
    \bottomrule
\end{tabularx}

\end{table}

 To evaluate the impact of the dependency graph on \textit{DRILS} performance, we compare the results of \textit{DRILS} with and without providing the dependency graph (\textit{DRILS-w/o-RDG}) as input to the \textit{Generator Agents}. In the latter setting, we remove the section of the main prompt that introduces the dependency graph and its relation types, along with the corresponding input graph. All other prompt sections and inputs remain unchanged. We use the same metrics as in RQ2 to analyze the results.

\subsubsection{\textbf{Overall Performance}}\label{subsec:RQ3OverallPerformance}
\autoref{tab:rq3} presents the results of \textit{DRILS-w/o-RDG} and \textit{DRILS}. Overall, \textit{DRILS} outperforms \textit{DRILS-w/o-RDG} on almost all metrics across all datasets. The average $F_1$ score increased from 76.7\% for \textit{DRILS-w/o-RDG} to 81\% for \textit{DRILS}, with \textit{DRILS} achieving $F_1$ scores above 80\% on two of the three datasets, whereas \textit{DRILS-w/o-RDG} remains below 80\% on all three datasets. The difference is more pronounced for recall, which increases from an average of 81.2\% for \textit{DRILS-w/o-RDG} to 87.1\% for \textit{DRILS}. Precision also improves when we incorporate the RDG, although the difference is less pronounced, with increases of 1.6 and 0.8 percentage points on D1 and D3, respectively. These results indicate that removing the dependency graph consistently reduces $F_1$ score, precision, and recall, suggesting that the dependency graph plays a critical role in directing the \textit{Generator Agents} to identify impacted requirements both more completely and more precisely. This may be because the RDG explicitly defines links between requirements, helping the agents identify additional requirements that may be impacted when a related requirement is considered impacted. This information is only explicitly defined in an RDG, making it a valuable input to \textit{DRILS}.

Including an RDG also positively impacts effectiveness, with effectiveness scores exceeding 90\% on both D1 and D2, a threshold that was not achieved by \textit{DRILS-w/o-RDG} on either dataset. Regarding cost, \textit{DRILS-w/o-RDG} achieved the lowest cost on D1 and D3, while costs were equal across both settings on D2. However, \textit{DRILS} was more robust, with cost deviations always at or below 0.05, compared with \textit{DRILS-w/o-RDG}, where cost deviations ranged from 0.06 to 0.10. Furthermore, the difference in cost was small across datasets, averaging 5.8\% for \textit{DRILS-w/o-RDG} and 6.0\% for \textit{DRILS}, a difference of only 0.2 percentage points. Therefore, incorporating an RDG improves effectiveness while keeping costs similar to \textit{DRILS-w/o-RDG}. Overall, given the consistent improvements across metrics, we consider RDG a valuable input for \textit{DRILS}.

\subsubsection{\textbf{Robustness Analysis}} Overall, the results are robust, with an average deviation of 0.038 across both settings. The deviation for \textit{DRILS-w/o-RDG} averages 0.05 compared with 0.026 for \textit{DRILS}. This reinforces the pattern observed in RQ2: providing additional context, in this case through an RDG, reduces variability across runs by giving the \textit{Generator Agents} more information to guide their responses. These results further highlight the importance of incorporating an RDG into the \textit{DRILS} architecture.

\begin{table*}
\centering
\caption{Generator Agents Stage Analysis - RDG}
\label{tab:generator-rdg}
\begin{tabularx}{\textwidth}{
    @{\hskip 0.5em}
    >{\centering\arraybackslash}m{0.08\textwidth}
    @{\hskip 0.5em}
    >{\raggedright\arraybackslash}m{0.15\textwidth}
    @{\hskip 0.5em}
    *{9}{>{\centering\arraybackslash}X}
    @{\hskip 0.5em}
}
    \toprule
    \multirow{3}{*}{\textbf{Category}} &
    \multirow{3}{*}{\textbf{Method}} &
    \multicolumn{3}{c}{\textbf{D1}} &
    \multicolumn{3}{c}{\textbf{D2}} &
    \multicolumn{3}{c}{\textbf{D3}} \\
    
    \cmidrule(lr){3-5}
    \cmidrule(lr){6-8}
    \cmidrule(lr){9-11}
    
    & &
    \textbf{R} & \textbf{P} & \textbf{F1} &
    \textbf{R} & \textbf{P} & \textbf{F1} & 
    \textbf{R} & \textbf{P} & \textbf{F1} \\
    
    \midrule
    \multirow{2}{*}{Agent1}  & DRILS-w/o-RDG & 76.4 & \textbf{23.1} & \textbf{35.3} & \textbf{66.7} & 16.2 & 26.2 & 77.8 & \textbf{26.7} & \textbf{40.8} \\
                         & DRILS         & \textbf{84.7} & 18.0 & 29.7 & 60.0 & \textbf{20.3} & \textbf{30.3} & \textbf{84.0} & 21.4 & 34.1 \\
\midrule
\multirow{2}{*}{Agent2} & DRILS-w/o-RDG  & \textbf{93.1} & 39.2 & 55.5 & \textbf{86.7} & 26.7 & 40.7 & 84.0 & 44.0 & 57.8 \\
                        & DRILS          & \textbf{93.1} & \textbf{39.4} & \textbf{57.0} & 81.7 & \textbf{36.4} & \textbf{51.9} & \textbf{85.2} & \textbf{46.0} & \textbf{59.8} \\
\midrule
\multirow{2}{*}{Union} & DRILS-w/o-RDG   & \textbf{95.8} & \textbf{24.1} & \textbf{38.5} & 86.7 & 17.2 & 28.7 & 86.4 & \textbf{24.3} & \textbf{37.9} \\
                       & DRILS           & \textbf{95.8} & 18.8 & 31.5 & \textbf{88.3} & \textbf{22.6} & \textbf{36.2} & \textbf{89.0} & 19.9 & 32.6 \\
    \bottomrule
\end{tabularx}

\end{table*}

\subsubsection{\textbf{Analysis of the Behavior of Generator Agents}}
\autoref{tab:generator-rdg} presents the performance metrics for each \textit{Generator Agent}, along with their union across all settings and datasets. The average $F_1$ scores for \textit{DRILS-w/o-RDG} are 34.1\% for Agent1, 51.3\% for Agent2, and 35.0\% for the union, while for \textit{DRILS}, they reach 31.3\%, 56.2\%, and 33.4\%, respectively. These results indicate that excluding the RDG yields higher $F_1$ scores for Agent1 and the union, while including it yields higher $F_1$ scores for Agent2. Notably, \textit{DRILS-w/o-RDG} achieves higher $F_1$ scores for Agent1 and the union because it performs better on D1 and D3. In contrast, \textit{DRILS} achieves the highest $F_1$ scores for Agent1, Agent2, and the union on D2. This may be because D2 has the fewest dependencies in its RDG (29), compared with 53 for D1 and 81 for D3, so including the RDG has less impact and yields results more similar to those without it. However, the differences in average $F_1$ score between the two settings are relatively small, with 2.8 percentage points in favor of \textit{DRILS-w/o-RDG} for Agent1, 4.9 percentage points in favor of \textit{DRILS} for Agent2, and 1.6 percentage points in favor of \textit{DRILS-w/o-RDG} for the union. Thus, since neither setting consistently outperforms the other and the differences in average $F_1$ score are relatively small, we need to examine other metrics more closely to identify the superior setting for the \textit{Generator Agents}.

For \textit{DRILS-w/o-RDG}, Agent1 achieves an average recall of 73.6\% and an average precision of 22.0\%, while Agent2 achieves 87.9\% and 36.6\%. For \textit{DRILS}, Agent1 achieves an average recall of 76.2\% and an average precision of 19.9\%, while Agent2 yields 86.7\% and 40.6\%. These results suggest that incorporating an RDG affects the two agents differently. For Agent1, it increases recall while reducing precision, suggesting that the agent identifies a broader set of candidate requirements. In contrast, Agent2 achieves higher precision at the cost of a slight decrease in recall, suggesting a more selective impact set identification.

Regarding the union, \textit{DRILS} achieves the highest union recall on all three datasets. This higher union recall suggests that incorporating an RDG helps the \textit{Generator Agents} produce a more complete set of candidate impacted requirements, providing the Judge Agent with a broader pool of candidates. This improvement may be attributable to the additional context provided by the RDG to the \textit{Generator Agents} (i.e., the links between requirements and the type of these links), as well as the potentially improved quality of the auxiliary information. One element of this auxiliary information is the confidence score. For \textit{DRILS-w/o-RDG}, the average confidence score for TPs and FPs is 4.69 and 3.65, respectively, compared with 4.73 and 3.60 for \textit{DRILS}. This suggests that incorporating the RDG is associated with higher confidence in correctly identified impacted requirements and lower confidence in false-positive predictions, which can help the Judge Agent distinguish between relevant and irrelevant candidates during filtering.

When an RDG is included, the \textit{Generator Agents} produced more diverse and complementary candidate sets. This is reflected in the differences between each agent's individual recall and their union recall, which reach 57.5 percentage points for \textit{DRILS}, compared with 53.1 percentage points for \textit{DRILS-w/o-RDG}.
One possible explanation is that incorporating the RDG encourages the \textit{Generator Agents} to explore different dependency paths when identifying impacted requirements, resulting in more diverse candidate sets, as reflected in \textit{DRILS}'s superior union recall. Finally, Consistent with the findings of RQ2, Agent2 outperforms Agent1 in terms of recall, precision, and $F_1$ score across all evaluated settings and datasets, reflecting its stronger overall performance in impact set identification.   

Overall, incorporating an RDG benefits the Judge Agent during filtering, potentially by improving the quality of the auxiliary information. Combined with the higher union recall observed at the Generator stage, these findings suggest that the RDG can benefit both the identification of candidate impacted requirements by the \textit{Generator Agents} and the subsequent filtering performed by the Judge Agent.

\begin{tcolorbox}[arc=1mm,width=\columnwidth,
                  top=0mm,left=0mm,  right=0mm, bottom=0mm,
                  boxrule=1pt, colback=blue!15!white,colframe=white]
\textbf{Answer to RQ3:} Integrating a requirements dependency graph improves overall recall, precision, $F_1$ score, and effectiveness. It benefits not only the \textit{Judge Agent} during validation but also helps the \textit{Generator Agents} produce impact sets with higher recall.
\end{tcolorbox} \subsection{RQ4: Comparison of \textit{DRILS} against Baseline Approaches}\label{sec:RQ4}

To address this research question, we compare \textit{DRILS} against four baselines, which were selected based on the criteria described in Section~\ref{sec:baselines}: (1) \textit{LSI}, an information retrieval approach, (2) \textit{SBERT}, a pre-trained embedding approach based on a language model, (3) \textit{LLM-Vec}, a pre-trained embedding approach based on a large language model, and (4) \textit{ProReFiCIA}, an LLM-based approach. The first three approaches (i.e., \textit{LSI}, \textit{SBERT}, and \textit{LLM-Vec}) generate textual embeddings of the change rationale and requirements, then identify the impact set by measuring their semantic similarity. For each of these approaches, we apply the two threshold selection strategies ($T_1$ and $T_2$) described in Section~\ref{sec:baselines} to derive the final impact set. Thus, our evaluation considers a total of seven baseline variants, six resulting from applying the two threshold selection strategies to each of the three approaches, and one corresponding to \textit{ProReFiCIA}. \autoref{tab:baseline} reports the results for all seven baseline variants, in addition to \textit{DRILS}, for the three datasets. The highest scores are highlighted in bold. As in the previous RQs, we use recall, precision, $F_1$ score, effectiveness, and cost as evaluation metrics.

\begin{table*}
\centering
\caption{Experimental Results of DRILS Against Baselines}
\label{tab:baseline}
\small
\begin{tabularx}{\textwidth}{
    @{\hskip 0.5em}
    >{\raggedright\arraybackslash}m{0.15\textwidth}
    @{\hskip 0.5em}
    *{15}{>{\centering\arraybackslash}X}
    @{\hskip 0.5em}
}
    \toprule
    \multirow{2}{*}{\textbf{Method}} &
    \multicolumn{5}{c}{\textbf{APA}} &
    \multicolumn{5}{c}{\textbf{TJA}} &
    \multicolumn{5}{c}{\textbf{ADB}} \\
    
    \cmidrule(lr){2-6}
    \cmidrule(lr){7-11}
    \cmidrule(lr){12-16}
    
    &
    \textbf{R} & \textbf{P} & \textbf{F1} & \textbf{Eff} & \textbf{Cost} &
    \textbf{R} & \textbf{P} & \textbf{F1} & \textbf{Eff} & \textbf{Cost} &
    \textbf{R} & \textbf{P} & \textbf{F1} & \textbf{Eff} & \textbf{Cost} \\
    
    \midrule
    \textbf{LSI$_{T1}$} & 37.5 & 10.7 & 16.6 & 52.5 & 35.3 & 45.0 & 50.0 & 47.4 & 63.9 & 64.8 & 37.0 & 27.0 & 31.2 & 47.5 & 43.1 \\
    \textbf{LSI$_{T2}$} & \textbf{91.6} & 12.9 & 22.7 & 87.5 & 15.9 & \textbf{85.0} & 27.9 & 41.9 & 91.6 & 30.3 & 77.8 & 18.8 & 30.2 & 70.8 & 18.8 \\
    \midrule
    \textbf{SBERT$_{T1}$} & 50.0 & 12.0 & 19.4 & 67.5 & 15.2 & 65.0 & 9.2 & 16.2 & 72.2 & 50.5 & 62.9 & 13.2 & 21.8 & 72.5 & 26.9 \\
    \textbf{SBERT$_{T2}$} & 83.3 & 10.3 & 18.4 & 91.6 & 29.4 & 80.0 & 19.0 & 30.8 & 88.9 & 30.1 & \textbf{96.3} & 20.6 & 33.9 & \textbf{98.3} & 27.2 \\
    \midrule
    \textbf{LLM-Vec$_{T1}$} & 58.3 & 6.0 & 10.8 & 77.5 & 35.6 & 80.0 & 10.7 & 18.8 & 88.9 & 54.8 & 48.2 & 12.0 & 19.3 & 66.7 & 29.4 \\
    \textbf{LLM-Vec$_{T2}$} & 83.3 & 9.5 & 17.1 & 91.6 & 31.7 & 65.0 & 18.1 & 28.3 & 83.3 & 26.5 & \textbf{96.3} & 17.8 & 30.0 & 95.8 & 30.4 \\
    \midrule
    \textbf{ProReFiCIA} & 83.3 & 44.5 & 58.0 & 91.6 & 6.8 & \textbf{85.0} & 48.8 & 62.0 & \textbf{91.6} & 12.5 & 84.7 & 30.7 & 45.1 & 84.7 & 13.5 \\
    \midrule
    \multirow{2}{*}{\textbf{DRILS}} & \textbf{91.6} & \textbf{64.9} & \textbf{75.9} & \textbf{94.9} & \textbf{5.1} & 83.3 & \textbf{86.2} & \textbf{84.8} & 82.5 & \textbf{6.8} & 86.4 & \textbf{77.8} & \textbf{81.9} & 91.7 & \textbf{6.2} \\
     & ${\scriptstyle \pm 0.00}$ & ${\scriptstyle \pm 0.05}$ & ${\scriptstyle \pm 0.03}$ &  ${\scriptstyle \pm 0.00}$ & ${\scriptstyle \pm 0.05}$ & ${\scriptstyle \pm 0.03}$ & ${\scriptstyle \pm 0.03}$ & ${\scriptstyle \pm 0.03}$ & ${\scriptstyle \pm 0.01}$ & ${\scriptstyle \pm 0.03}$ & ${\scriptstyle \pm 0.02}$ & ${\scriptstyle \pm 0.04}$ & ${\scriptstyle \pm 0.03}$ & ${\scriptstyle \pm 0.01}$ & ${\scriptstyle \pm 0.03}$ \\
    \bottomrule
\end{tabularx}

\end{table*}

\subsubsection{\textbf{Comparison of Threshold Selection Strategies}} 

For \textit{LSI}, \textit{SBERT}, and \textit{LLM-Vec}, we experimented with two threshold selection strategies. For \textit{LSI}, $T_1$ achieves a superior $F_1$ score for two out of three datasets. However, the average $F_1$ score is similar between the two strategies, reaching 31.7\% for $T_1$ compared to 31.6\% for $T_1$. For \textit{SBERT} and \textit{LLM-Vec}, $T_1$ yields a superior $F_1$ score on two and three of the three datasets, respectively. Moreover, $T_2$ achieves average $F_1$ scores of 27.7\% and 25.1\% for \textit{SBERT} and \textit{LLM-Vec}, respectively, compared to 19.1\% and 16.3\% for $T_1$. Although the differences in $F_1$ scores are relatively marginal, $T_1$ provides substantially higher recall for almost all experiments. For LSI, the average recall increases from 39.8\% to 84.8\%, while for \textit{SBERT} and \textit{LLM-Vec}, it increases from 59.3\% to 86.5\% and from 62.1\% to 81.5\%, respectively. This corresponds to relative gains of 113.1\%, 45.9\%, and 31.2\%, respectively. This substantial difference in recall can be attributed to setting $T_2$ below the average of the higher scores, which generally lowers the threshold and thus includes more requirements in the predicted impact set, increasing the chance of capturing true positives. Nonetheless, the $F_1$ score for each approach remains below 50\%, indicating that semantic similarity alone is insufficient to accurately identify the impact set.

In terms of effectiveness, $T_2$ outperformed $T_1$ in 8 of 9 configurations, while its cost was lower in 6 of 9 experiments. Across the three approaches and three datasets, $T_2$ achieves an average effectiveness of 88.8\% and an average cost of 26.7\%, compared to 69.6\% and 39.5\% for $T_1$, respectively. These results indicate that $T_2$ not only provides higher $F_1$ scores and substantially higher recall, but also achieves greater effectiveness with more cost savings, making it a superior threshold-selection strategy to $T_1$.

\subsubsection{\textbf{Comparison of Baseline Approaches}} 

Among all baselines, \textit{ProReFiCIA} achieves the highest $F_1$ score across the three datasets. Compared with the other three baselines using the $T_2$ threshold selection strategy, it achieves an average $F_1$ score of 55.0\%, versus 31.6\%, 27.7\%, and 25.1\% for \textit{LSI}, \textit{SBERT}, and \textit{LLM-Vec}, respectively. This represents a substantial improvement, with \textit{ProReFiCIA} achieving nearly twice the average $F_1$ score of the three semantic similarity-based approaches. This difference is primarily caused by the significant difference in precision between \textit{ProReFiCIA} and the other baselines. \textit{ProReFiCIA} achieves an average precision of 41.33\%, compared to 19.9\%, 16.6\%, and 15.1\%, for \textit{LSI}, \textit{SBERT}, and LLM-Vaec, respectively. This can be attributed to its more balanced retrieval of true positives and false positives. Unlike the semantic similarity approaches, which tend to either miss relevant requirements or include too many irrelevant ones, \textit{ProReFiCIA} produces a more precise impact set. After \textit{ProReFiCIA}, and by a considerable margin, the next highest-performing baseline is \textit{LSI}$_{T2}$, as indicated by the average $F_1$ scores reported above. This suggests that richer semantic representations, without any further fine-tuning, do not necessarily translate to better retrieval performance in the CIA task. Among the remaining two pre-trained embedding models, \textit{SBERT}$_{T2}$ outperforms \textit{LLM-Vec}$_{T2}$, with the latter achieving the lowest $F_1$ score overall.

In terms of cost, the same pattern holds: \textit{ProReFiCIA} has the lowest cost at 10.9\%, followed by \textit{LSI} at 21.6\%, \textit{SBERT} at 28.9\%, and \textit{LLM-Vec} at 29.5\%. 
Regarding effectiveness, a different pattern emerges, with approaches based on pre-trained embeddings achieving the highest effectiveness. SBERT$_{T2}$ achieves the highest average effectiveness at 92.9\%, followed by \textit{LLM-Vec}${T2}$ at 90.2\%, \textit{ProReFiCIA} at 89.3\%, and \textit{LSI}$_{T2}$ at 89.3\%. These results indicate that although pre-trained embedding models achieve lower $F_1$ scores than other baselines, they can still provide highly effective impact-set retrieval. This may be because they capture richer semantic relationships between requirements than a simpler method such as \textit{LSI}. Nevertheless, \textit{ProReFiCIA} achieves the highest $F_1$ score and lowest cost, making it the strongest overall baseline.

\subsubsection{\textbf{Comparison of \textit{DRILS} with Baselines}} As \autoref{tab:baseline} shows, \textit{DRILS} achieves the best performance on most metrics across the three datasets, outperforming the baselines in 11 out of 15 metric-dataset combinations. \textit{DRILS} achieves an average recall of 87.1\%, an average precision of 76.3\%, and an average $F_1$ score of 80.9\% across the three datasets. Compared to the best-performing baseline, \textit{ProReFiCIA}, this corresponds to relative gains of 30.86\%, 36.77\%, and 82.0\%. This difference is primarily due to the substantially higher precision achieved by \textit{DRILS}. While all four baselines (using the $T_2$ threshold selection strategy) achieve similar recall scores and even outperform \textit{DRILS} on the ADB dataset, the difference is much more significant when considering precision. In fact, compared to \textit{ProReFiCIA}, \textit{DRILS} increases precision from 44.5\% to 64.9\% for APA, from 48.8\% to 86.2\% for TJA, and from 30.7\% to 77.8\% for ADB. These correspond to relative gains of 41.1\%, 83.4\%, and 149.8\%, respectively. Thus, \textit{DRILS} substantially improves precision while maintaining comparable recall, resulting in a considerably better balance between precision and recall. The difference is even more pronounced when considering other baselines. Precision is as low as 6.0\% with \textit{LLM-Vec}${T1}$ for APA, 9.2\% with \textit{SBERT}${T1}$ for TJA, and 12.0\% with \textit{LLM-Vec}$_{T1}$ for ADB. Overall, \textit{DRILS} achieves the highest $F_1$ scores across the evaluated approaches, primarily due to its substantially higher precision, while maintaining high recall. These results demonstrate the advantage of \textit{DRILS} in producing better impact sets compared to the baseline approaches.

In terms of effectiveness and cost, \textit{DRILS} achieves the highest effectiveness on two of the three datasets and the lowest cost on all three. It yields an average effectiveness of 89.7\% and an average cost of 6.0\% across the three datasets. This means that on average, \textit{DRILS} reviews 6.0\% of the requirements list per change request while maintaining an effectiveness of approximately 90\%. These results indicate that \textit{DRILS} not only avoids overwhelming analysts with irrelevant requirements, but also retains the ability to identify the vast majority of truly impacted requirements. Although \textit{ProReFiCIA} was specifically designed to reduce the cost of impact analysis, \textit{DRILS} achieves substantially lower costs. In particular, \textit{DRILS} reduces the cost by nearly half compared to \textit{ProReFiCIA} for both the TJA and ADB datasets. The difference is even more pronounced than for other baselines, particularly \textit{LSI}$_{T1}$, which achieves the lowest effectiveness and highest cost across all three datasets. Specifically, \textit{LSI}$_{T1}$ achieves an average effectiveness of only 54.0\% at an average cost of 47.7\%, compared to 89.7\% effectiveness and 6.0\% cost for \textit{DRILS}. Overall, DRILS achieves a high effectiveness while providing significant cost savings compared to the baselines. Combined with its superior precision, these results demonstrate \textit{DRILS}'s ability to reduce false positives, thus reducing the effort required by requirements engineers to review the predicted impact set.

To assess whether the performance difference between \textit{DRILS} and \textit{ProReFiCIA}, the highest-performing baseline, is statistically significant, we conducted Fisher's Exact test. Specifically, Fisher's exact test was used to determine whether the observed differences in precision and recall arise from statistically significant differences in the TP/FP and TP/FN proportions. 
Regarding recall, the p-values for all three datasets are above the significance threshold of 0.05. Therefore, there is no statistically significant difference in recall between the two approaches. In contrast, the p-values for precision across the three datasets are 2e-2, 3.3e-3, and 1e-4 for APA, TJA, and ADB, respectively. Since these values are below the significance threshold, the differences in precision are statistically significant for all three datasets. These results indicate that \textit{DRILS}'s superior $F_1$ scores are driven by higher precision rather than a statistically significant improvement in recall. Together, these results, along with \textit{DRILS}'s highest $F_1$ score, high effectiveness, and lowest cost, establish \textit{DRILS} as the superior approach among the evaluated baselines.

\begin{tcolorbox}[arc=1mm,width=\columnwidth,
                  top=0mm,left=0mm,  right=0mm, bottom=0mm,
                  boxrule=1pt, colback=blue!15!white,colframe=white]
\textbf{Answer to RQ4:} \textit{DRILS} consistently achieves the highest $F_1$ score among the evaluated baselines with an average of 81\% across datasets. This improvement is primarily driven by a statistically significant increase in precision. Additionally, \textit{DRILS} achieves high effectiveness at the lowest cost. These results demonstrate \textit{DRILS}'s ability to accurately identify the impact set while reducing false positives, thus substantially reducing the effort required by requirements engineers to review the predicted impact set.
\end{tcolorbox}

  \section{Discussion}\label{sec:discussion}
\textit{DRILS} comprises multiple steps, inputs, and outputs. Through extensive experimental evaluations, we demonstrated the usefulness of each of these elements. We further demonstrated that \textit{DRILS} consistently outperformed state-of-the-art (SOTA) baselines in terms of recall, $F_1$ score, effectiveness, and cost, while achieving substantially higher precision.

\paragraph{\textbf{Auxiliary Information for Improved Impact Set Validation}} Our evaluations demonstrate that auxiliary information provided by the generator agents contributes to the \textit{Judge Agent}’s decision-making. Including rationales improved precision, recall, and $F_1$ score, with the most pronounced gains among the evaluated types of auxiliary information. Including change locations primarily improved recall, while including Confidence Scores improved recall, precision, and $F_1$ score, albeit with slight differences. Each type of information gives the Judge Agent a different perspective: rationales explain the Generator Agents' reasoning, change rationales provide more detail about the proposed change to the impacted requirement, and confidence scores indicate the agents' confidence in their decisions. These types of information complement each other, as demonstrated by the more pronounced improvements across all evaluation metrics when they were included simultaneously.

These findings highlight the importance of providing agents in LLM-as-a-judge setups with auxiliary information that offers transparency into the reasoning and confidence of preceding agents. Such information can help the Judge Agent make more informed decisions for CIA.

\paragraph{\textbf{RE-RAG over Chunk-Based RAG: Retrieval Quality over Quantity}} The experimental results show that RE-RAG improves the performance of \textit{DRILS} compared with both chunk-based RAG and no RAG. RE-RAG provides superior overall performance, particularly in recall, producing more complete impact sets. Unconstrained context from chunk-based RAG yields results similar to those without RAG, potentially due to introduced noise, highlighting the importance of concise, structured context. RE-RAG particularly benefits the Judge Agent during impact set validation, rather than the Generator Agents during impact set identification. This may be explained by the higher quality of the auxiliary information produced by the Generator Agents when they receive concise and structured domain context as input.

These findings highlight the importance of concise and structured context, as unconstrained context yields results similar to those obtained without context, potentially due to the noise introduced. 

\paragraph{\textbf{Clustering over Batching: Semantic Grouping over Size-Based Grouping}} The empirical results show that clustering the input requirements in \textit{DRILS} yields higher recall, and thus a more complete impact set, compared with fixed-size batching and processing the entire requirements list at once. This may be because clustering reduces the amount of information the agents must process simultaneously while preserving the context of semantically similar requirements. This allows them to focus more effectively on requirements relevant to the change request while mitigating the lost-in-the-middle effect. Clustering particularly benefits impact-set validation by the \textit{Judge Agent}, rather than impact-set generation by the \textit{Generator Agents}. 
This may be due to higher-quality auxiliary information, as evidenced by higher confidence scores, which better differentiate true positives (TPs) from false positives (FPs).

These findings highlight the importance of considering semantic similarity when dividing requirements for processing, particularly in complex systems with a large number of requirements, as semantically similar requirements provide additional relevant context for CIA.

\paragraph{\textbf{Requirement Dependency Graphs: Valuable Inputs for LLM-based CIA}} The empirical results show that integrating a requirement dependency graph as additional knowledge improves \textit{DRILS}'s overall recall, precision, $F_1$ score, and effectiveness. It benefits both impact set identification and validation. This may be attributed to the additional context provided by the RDG beyond the domain context, specifically the information regarding relationships between requirements. This context helps the agents identify additional requirements that may be impacted when a related requirement is deemed impacted. It can also support generating higher-quality, better-grounded auxiliary information, as shown by confidence scores that better distinguish true positives from false positives. 

These findings highlight the value of incorporating requirement dependencies into LLM-based change impact analysis, supporting both impact set identification and validation. 

\paragraph{\textbf{\textit{DRILS}: Improved Performance Compared with Baselines}} The empirical results show that \textit{DRILS} outperforms multiple baselines across three categories: information retrieval, pre-trained embeddings, and LLMs. \textit{DRILS} achieves higher recall, precision, $F_1$ scores, and effectiveness, while incurring a lower cost. Although \textit{DRILS} achieves high recall, averaging 87\%, its main strength lies in precision, which is nearly twice that of the best-performing baseline. This substantially reduces the effort required by analysts to review the final impact set. Combined with the rationales explaining the selection of impacted requirements, \textit{DRILS} facilitate the subsequent manual validation performed by analysts.

These findings highlight the potential of LLMs for CIA, particularly when advanced techniques are employed to provide grounded context and leverage multi-agent architectures. Although manual validation remains necessary, LLM-based CIA, particularly \textit{DRILS}, can substantially facilitate the task and automate a large part of the analysis, thereby reducing the manual verification effort required from analysts.  \section{Threats to Validity}\label{sec:threats}
\sectopic{Internal Validity.}
One internal threat to validity concerns the design choices made for \textit{DRILS}, particularly the integration of various inputs, outputs, and steps within its architecture. To mitigate this threat, we introduced these elements systematically and assessed their contributions through the extensive ablation studies conducted in RQ1, RQ2, and RQ3. Another threat is related to these ablation studies: whether the difference in performance can be attributed to the specific elements under investigation or due to external factors. To address this, we conducted controlled experiments in which all other inputs, steps, and prompts remained unchanged, except for the element being evaluated.
Another threat arises from the stochastic nature of LLMs, which can lead to variations in results across runs. To mitigate this, we set the temperature to 0 to minimize randomness and re-ran all experiments three times. We report the average results and analyze the standard deviation, which was negligible, suggesting limited variability across runs and demonstrating the robustness of the experimental results. Finally, the observed performance differences between \textit{DRILS} and the baseline approaches may be due to random variation. To mitigate this, we conducted Fisher’s exact test to assess statistical significance, providing stronger evidence that the reported improvements reflect genuine performance differences.

\sectopic{External Validity.} The most critical threat to external validity concerns the generalizability of \textit{DRILS} across new datasets. In this work, we evaluated \textit{DRILS} on three different automotive datasets, covering a variety of systems with different numbers of requirements and dependencies. While the number of requirements in these datasets may not reflect the scale of larger industrial systems, we explicitly considered this challenge in the design of \textit{DRILS}, particularly through the clustering mechanism introduced in Section~\ref{sec:RQ2}, which was designed to maintain effectiveness as the number of requirements grows. Nevertheless, evaluating \textit{DRILS} on datasets from other domains remains an important direction for future work. Another threat concerns the three LLMs used in \textit{DRILS}: GPT, Claude, and Gemini. Although these models were selected based on their demonstrated effectiveness and use in requirements engineering research, other LLMs may yield different results. Nevertheless, we expect the main findings to generalize to LLMs with comparable capabilities. As future work, we plan to evaluate additional proprietary and open-source LLMs to further assess the generalizability of our findings. 

\sectopic{Construct Validity.} A threat to construct validity concerns our manual annotation of the ground truth impact sets for our datasets, which may introduce subjectivity and bias. To mitigate this, we followed a systematic, iterative annotation process, described in Section~\ref{sec:dataset}, in which two researchers independently identified the impact sets and cross-validated their results through discussion until they reached consensus. Another threat concerns the replication of the baseline approaches. To mitigate this threat, we followed the implementation guidelines provided in the corresponding literature and official documentation, as described in Section~\ref{sec:implementation}.

\sectopic{Conclusion Validity.} Another important threat concerns the reproducibility of our results. To mitigate this, we report all implementation details with sufficient precision. We also make all experimental artifacts publicly available~\cite{annex}, enabling independent verification and replication of our findings under identical conditions.  \section{Related Works}\label{sec:related-works}

Arnold and Bohner~\cite{arnold1996software, bohner1996impact} laid the foundations of change impact analysis in 1993, defining it as a strategic process for evaluating the ripple effects of a modification and determining precisely which system artifacts require updates. In requirements engineering, CIA is particularly challenging because requirements are often complex, interdependent, and ambiguous due to the use of NL.
To address these challenges, various CIA approaches were proposed in the literature. These approaches can be broadly grouped into four categories: model-based approaches~\cite{aryani2009change,nejati2016automated,kchaou2017uml,alkaf2019automated}, requirement dependency-based approaches~\cite{goknil2008change,goknil2011semantics,nonsiri2013model,goknil2014change,rahim2026ontology}, NLP-based approaches~\cite{arora2015change,arora2015narcia}, and LLM-based approaches~\cite{rusdianto2024innovative,etezadi2025llm}. 

\subsection{Model-Based Approaches}

Model-based approaches represent software artifacts, including requirements, using formal representations, which are analyzed using techniques such as static slicing and graph traversal to support CIA. Specifically, they leverage the structure and semantics of formal models to identify change impact across software artifacts.
Early foundational work in this category was proposed by Aryani~\textit{et al.}~\cite{aryani2009change}, a methodology for analyzing change propagation using domain-level behavioral models. These models capture the structural interactions between system components and are constructed by representing components and their behavioral dependencies at a conceptual level. Change propagation then traverses these behavioral dependencies, following how a change in one component influences others through their interactions, making the propagation mechanism primarily structural and topology-driven. Nejati~\textit{et al.}~\cite{nejati2016automated} also addressed cross-artifact CIA, specifically between SysML requirements and design models. The approach first applies a static slicing algorithm to extract the set of potentially impacted model elements. The identified candidate elements are then ranked using NLP applied to their textual content and manually inspected by analysts. This approach combines structural slicing for propagation with semantic similarity scoring for prioritization, making it a notable hybrid of structural and NLP-based analysis. 

Kchaou~\textit{et al.}~\cite{kchaou2017uml} support CIA across UML design artifacts by leveraging structural dependencies within and between UML diagrams. They employ an information retrieval (IR) technique that measures textual similarity between natural language descriptions associated with model elements. Lastly, Alkaf~\textit{et al.}~\cite{alkaf2019automated} proposed an automated CIA approach for requirements modeled using the User Requirements Notation (URN), a formalism that combines Use Case Maps (UCM) for expressing scenarios and Goal-oriented Requirement Language (GRL) for capturing the goals of actors and their relationships. Given a change to a URN construct, the approach uses forward slicing and URN traceability links to identify potentially impacted constructs within the selected view and across related GRL/UCM views.

\subsection{Requirement Dependency-Based Approaches}

Dependency-based techniques focus on explicitly defined relationships among requirements, such as refinement, conflict, or derivation, and use these relationships to trace how a change propagates from one requirement to others~\cite{goknil2008change,goknil2011semantics}. They operate primarily at the requirement level, propagating changes through typed dependency links. For instance, G{\"o}knil~\textit{et al.}~\cite{goknil2008change} formalized a subset of requirement relationships in first-order logic. Change propagation is performed through deductive rule-based inference, deriving the set of impacted requirements by applying logical rules to the formalized relationship semantics. Only a textual explanation of inconsistency propagation was provided, and no fully automated tool was developed. Similarly, Goknil~\textit{et al.}~\cite{goknil2011semantics} proposed an approach that relies on first-order logic formalization of requirement relations and dependencies. They implemented a reasoning tool based on an OWL ontology that represents the requirement-dependency graph. Change propagation and consistency checking are performed through rule-based logical reasoning, leveraging OWL description logic to derive impacted requirements from formally encoded relationship semantics.

Multiple existing approaches are hybrid, combining both modeling and requirement dependencies for CIA. For instance, Nonsiri~\textit{et al.}~\cite{nonsiri2013model} proposed a model-based approach for analyzing change propagation among requirements by automatically transforming SysML requirement diagrams into OWL ontologies. The transformation pipeline maps requirement elements and their relationships to ontological concepts and properties, constructing the dependency model without manual ontology authoring. Change propagation is then performed through rule-based reasoning.
Goknil~\textit{et al.}~\cite{goknil2014change} defined a meta-model governing the structure and semantics of requirement models, representing requirements and their dependencies as instances that conform to this meta-model. Change propagation is performed through a combination of model traversal and rule application over the meta-model structure. However, the approach requires analysts to align their requirement specifications with the prescribed meta-model.
Finally, Rahim~\textit{et al.}~\cite{rahim2026ontology} proposed an approach that automatically transforms SysML requirement diagrams into OWL ontologies by mapping SysML elements and their relationships to ontological concepts and properties to construct the dependency model. Once the ontology is built, change propagation is performed by an OWL reasoning engine executing predefined semantic change impact rules that encode how relationship types propagate a change from one element to others.

\subsection{NLP-based Approaches}

Natural Language Processing (NLP) approaches primarily focus on the textual and linguistic characteristics of requirements and change requests. NLP techniques support CIA by leveraging the lexical and syntactic features of requirements and change requests to identify potentially impacted requirements. For example, Arora~\textit{et al.}~\cite{arora2015change,arora2015narcia} proposed an NLP-based approach that analyzes how changes affect natural language requirements. Their approach combined text chunking to extract noun phrases and verb phrases from change rationales and candidate requirements with analyst-defined change propagation conditions. These conditions used Boolean logic over phrases or their synonyms and enabled phrase-level, context-aware impact scoring that reflects the likelihood that each requirement is affected.

\subsection{LLM-Based Approaches}

More recently, researchers have explored LLM-based approaches to CIA. These approaches leverage the natural language understanding and reasoning capabilities of LLMs to identify potentially impacted requirements. For instance, Rusdianto~\textit{et al.}~\cite{rusdianto2024innovative} proposed an LLM-based framework that integrates GPT-4 with RAG to identify potentially impacted requirements and analyze the functional, business, technical, and stakeholder implications of requirement changes. 
Another example is ProReFiCIA~\cite{etezadi2025llm}, an LLM-based approach that follows a structured pipeline consisting of prompting, refinement, and natural language inference (NLI)-based filtering stages. This pipeline aims to accurately identify impacted requirements while maintaining high recall and reducing both false positives and the manual review effort associated with them.

\subsection{Limitations and \textit{DRILS} Contribution}

The aforementioned approaches exhibit several limitations. Model-based approaches rely on formal representations that support systematic and accurate change impact analysis. However, they require requirements to conform to a specific modeling formalism (e.g., URN and SysML), limiting their applicability across different development contexts~\cite{aryani2009change,nejati2016automated,rahim2026ontology,alkaf2019automated,nonsiri2013model,goknil2014change}. Furthermore, most of of these approaches address change propagation across software artifacts~\cite{aryani2009change,nejati2016automated,rahim2026ontology}, rather than requirement change impact identification, which is the focus of this study. Requirement dependency-based approaches leverage dependencies between requirements to provide meaningful context for CIA. However, these structural dependencies alone are generally insufficient to capture the full semantic implications of a change request, particularly when relevant relationships are implicit or informally expressed~\cite{goknil2008change,goknil2011semantics,nonsiri2013model,goknil2014change}. Moreover, in industrial contexts, requirement dependencies are often incomplete and sparsely defined, making approaches that rely solely on them less applicable. 
NLP-based techniques offer greater flexibility in processing natural-language requirements but often rely on shallow linguistic features that fail to capture deeper semantic relationships. They also require substantial human involvement, limiting their scalability and application~\cite{arora2015change,arora2015narcia}. 
Finally, recent LLM-based approaches rely on simple prompting and basic RAG, without retrieving richer contextual information or leveraging requirement dependencies, potentially overlooking impacted requirements~\cite{rusdianto2024innovative,etezadi2025llm}. 

\textit{DRILS} addresses these gaps through three key features. First, it adopts an agentic framework that uses multiple LLM models, enabling enhanced reasoning over complex change requests. Second, it leverages domain and technical knowledge embedded within the SRS document through a tailored RAG technique (RE-RAG), retrieving contextually relevant information more effectively than conventional approaches. Third, it incorporates dependency graph information as a complementary source of evidence alongside semantic reasoning, rather than treating structural dependencies as the sole basis for impact analysis. Together, these features make \textit{DRILS} a promising solution for requirements change impact analysis.

 \section{Conclusion}\label{sec:conclusion}
This paper presented \textit{DRILS}, an agentic LLM-based framework for requirements change impact analysis that leverages a requirements dependency graph. \textit{DRILS} incorporates RE-RAG, a tailored retrieval-augmented generation technique, to enhance accuracy. Two distinct Generator Agents are employed for impact set identification to introduce diversity of reasoning within the framework. They are each guided by a dependency graph that captures inter-requirement relationships to improve the completeness of their generated impact sets. A Judge Agent filters out false positives from the generated results to improve the precision of impact set identification. 

We evaluated \textit{DRILS} on three datasets drawn from the automotive domain.
The experiments demonstrate that \textit{DRILS} significantly outperforms state-of-the-art baselines, particularly in precision, while maintaining high recall. \textit{DRILS} achieves an average $F_1$ score of 81\% across all three datasets, with individual scores of 76\%, 85\%, and 82\%. 
The average recall and precision of \textit{DRILS} are 87\% and 76\%, respectively, indicating that the framework is capable of identifying the vast majority of truly impacted requirements while maintaining strong precision. Moreover, the ablation experiments revealed that removing RE-RAG, clustering, or the Requirement Dependency Graph reduced recall and $F_1$ score. Removing the auxiliary information produced by the generator agents also reduced recall, with a more pronounced relative decrease of 18\% when all auxiliary information was removed simultaneously. 
These findings underscore the essential contribution of each component of \textit{DRILS} to its overall performance.

As future work, we plan to investigate alternative multi-agent configurations, including agent debate and the use of more than two generator agents, to assess their impact on \textit{DRILS}' performance. We also aim 
to evaluate the generalizability of \textit{DRILS} beyond the automotive domain using diverse datasets spanning a broader range of application domains and requirement characteristics. Furthermore, we plan to assess its performance with different proprietary and open-source LLMs to investigate the extent to which the results depend on the underlying models. 
Finally, we aim to extend \textit{DRILS} towards end-to-end CIA across multiple software artifacts, including design, development, and testing artifacts.

\begin{acks}

This work was supported by a research grant from General Motors, as well as the Canada Research Chair and Discovery Grant programs of the Natural Sciences and Engineering Research Council of Canada (NSERC). Lionel C. Briand’s contribution was partially funded by the Research Ireland grant 13/RC/209.

\end{acks}

\balance
\bibliographystyle{ACM-Reference-Format}
\bibliography{paper}

\end{document}